\documentclass[english, a4paper, 11pt, DIV=12, numbers=noenddot]{scrartcl}
\usepackage{blindtext}
\usepackage{amsmath}
\usepackage{amssymb}
\usepackage{bbm}
\usepackage{slashed}
\usepackage{cite}
\usepackage{setspace}
\usepackage[bottom]{footmisc}
\usepackage{hyperref}
\usepackage{graphicx}
\usepackage[labelfont=bf]{caption}
\usepackage{subcaption}
\usepackage{braket}
\usepackage{chngcntr}
\usepackage{booktabs}
\usepackage{color, colortbl}
\usepackage[dvipsnames]{xcolor}
\usepackage{multirow}

\numberwithin{equation}{section} 
\counterwithin{figure}{section} 
\counterwithin{table}{section} 
\hypersetup{linktocpage,
    colorlinks,
    linkcolor={red!70!black},
    citecolor={green!50!blue},
    urlcolor={blue!50!black}
}
\newcommand*{\GtrSim}{\smallrel\gtrsim}
\newcommand*{\LessSim}{\smallrel\lesssim}

\makeatletter
\newcommand*{\smallrel}[2][.8]{%
  \mathrel{\mathpalette{\smallrel@{#1}}{#2}}%
}
\newcommand*{\smallrel@}[3]{%
  \sbox0{$#2\vcenter{}$}%
  \dimen@=\ht0 %
  \raise\dimen@\hbox{%
    \scalebox{#1}{%
      \raise-\dimen@\hbox{$#2#3\m@th$}%
    }%
  }%
}
\makeatother

\newcommand{\stkout}[1]{\ifmmode\text{\sout{\ensuremath{#1}}}\else\sout{#1}\fi}

\begin{document}
\begin{titlepage}
\begin{flushright}
TTP22-063\\
P3H-22-104
\end{flushright}
\vskip2cm
\begin{center}
{\LARGE \bfseries Lepton-Flavoured Scalar Dark Matter \\in Dark Minimal Flavour Violation}
\vskip1.0cm
{\large Harun Acaro\u{g}lu$^{a,b}$, Prateek Agrawal$^b$, Monika Blanke$^{a,c}$}
\vskip0.5cm
 \textit{$^a$Institut f\"ur Theoretische Teilchenphysik,
  Karlsruhe Institute of Technology, \\
Engesserstra\ss e 7,
  D-76128 Karlsruhe, Germany}
 \vspace{3mm}\\
  \textit{$^b$Rudolf Peierls Centre for Theoretical Physics, University of Oxford,\\
  Parks Road, Oxford OX1 3PU, United Kingdom}
  \vspace{3mm}\\
  \textit{$^c$Institut f\"ur Astroteilchenphysik, Karlsruhe Institute of Technology,\\
  Hermann-von-Helmholtz-Platz 1, D-76344 Eggenstein-Leopoldshafen, Germany}

\vskip1cm


\vskip1cm

{\large \bfseries Abstract\\[10pt]} \parbox[t]{.9\textwidth}{
We study a simplified model of lepton-flavoured complex scalar dark matter set up in the Dark Minimal Flavour Violation framework. In this model the Standard Model is extended by a scalar dark matter flavour triplet and a charged fermionic mediator, through which dark matter couples to the right-handed charged leptons of the Standard Model. This interaction is parameterized by a new $3\times 3$ coupling matrix $\lambda$. Consistent with the field content of the model, also the Standard Model's approximate flavour symmetry is extended to include an additional global $U(3)$ associated with the dark matter flavour triplet. In addition to the Standard Model Yukawa couplings, the new coupling matrix $\lambda$ is assumed to constitute the only source that violates this extended symmetry. We analyse the parameter space of this model by investigating constraints from collider searches, lepton flavour violating decays, the observed dark matter relic density, and direct as well as indirect dark matter detection experiments. By performing a combined analysis of all constraints we find that restrictions from lepton flavour violating decays, the observed relic density and dark matter nucleon scattering are dominant. The combination of the latter two renders limits from collider searches irrelevant while indirect detection constraints are weak due to a $p$-wave suppression of the annihilation rate. We conclude that lepton-flavoured scalar dark matter has a rich phenomenology and is a viable dark matter candidate.}

\end{center}
\end{titlepage}

\tableofcontents 
\newpage

\section{Introduction}
\label{sec:intro}
Although the existence of dark matter (DM) is backed up by solid evidence \cite{Bertone:2004pz,Planck:2018vyg}, its particle physics properties still remain unknown. There are numerous dark matter models and mapping this vast theory space to experimental signatures is a challenging problem. Even within the paradigm of weakly interacting massive particles (WIMPs), it is often useful to appeal to a simplified model framework that categorizes the signals from large classes of models.\par
One such class of models is the idea of flavoured dark matter (FDM) \cite{Kile:2011mn,Kamenik:2011nb,Batell:2011tc,Agrawal:2011ze,Batell:2013zwa,Kile:2013ola,Kile:2014jea, Lopez-Honorez:2013wla,Kumar:2013hfa, Zhang:2012da}. These models are motivated by the observation that stringent flavour constraints imply that DM at the weak scale cannot have arbitrary interactions with Standard Model (SM) matter, but must have a very specific flavour structure. In these models, in analogy to the SM matter content, DM transforms under the fundamental representation of an associated dark flavour symmetry -- i.e.\@ DM is flavoured and comes in multiple (usually three) generations.\par 	
While FDM models exhibit some phenomenological advantages like additional DM annihilation channels due to the increased number of dark particles, they also come at the cost of strong constraints from flavour experiments. Hence, early studies of such models have limited their analysis to the minimal flavour violation (MFV) paradigm \cite{Buras:2000dm,DAmbrosio:2002vsn,Buras:2003jf,Chivukula:1987py,Hall:1990ac,Cirigliano:2005ck}, where it is assumed that the only sources of flavour violation are the SM Yukawa couplings. A consequence of this premise is that the coupling matrix of DM to SM matter has to be expressed in terms of the latter, which in turn yields a highly restricted flavour structure.\par 
A more general approach to study FDM models is to go beyond the MFV hypothesis and allow the new coupling matrix to constitute an additional source of flavour and CP violation. One framework for the analysis of this class of models is the Dark Minimal Flavour Violation (DMFV) framework which was introduced in \cite{Agrawal:2014aoa}. Here, the FDM ansatz is extended to allow for a non-trivial flavour structure of the new coupling matrix $\lambda$ that governs the interaction of DM with SM fermions.\par 
Even within the DMFV framework our ignorance about fundamental properties of DM leaves us with many options in terms of model building. Apart from the choice of the particle nature of DM and its corresponding mediator particle, one is also left with a large variety of options in terms of the SM fields that DM interacts with. Here, the DMFV framework allows for couplings to all fermion types of the SM. The cases of DM being a Dirac fermion coupling to the various SM quark fields have already been studied intensively \cite{Agrawal:2014aoa, Blanke:2017tnb, Blanke:2017fum, Jubb:2017rhm}, and its coupling to the right-handed charged leptons has been investigated in \cite{Chen:2015jkt}. More recently a study of a Majorana fermionic DMFV model has been presented in \cite{Acaroglu:2021qae}. In the present study we instead assume DM to be a complex scalar that couples to the right-handed charged leptons.\par 
Lepton-flavoured, lepton-portal or leptophilic DM again corresponds to a class of DM models with specific features \cite{Bai:2014osa, Lee:2014rba, Hamze:2014wca, Kawamura:2020qxo}. One significant property of such models in general is the lack of tree-level contributions to DM-nucleon scattering, leading to a suppression of the signal in direct detection experiments. Further, lepton-flavoured DM models are less constrained by LHC searches, as the corresponding mediator particle of leptophilic DM can only be produced by a Drell-Yan process and hence suffers from an $s$-channel suppression combined with the smallness of the electroweak coupling. These features together ameliorate the tension between the WIMP paradigm and the absence of a signal in such experiments. However, choosing the DM to interact with leptons on the other hand generally comes at the cost of stronger indirect detection constraints due to the direct coupling to electrons and positrons.\par 
We begin our analysis by briefly reviewing the DMFV framework and setting up the above-mentioned simplified model of lepton-flavoured scalar DM. We then study its phenomenology by subsequently investigating constraints from collider, flavour, cosmology and direct as well as indirect detection experiments. To provide a global picture we complete our study by performing a combined analysis in which we demand that all constraints are satisfied at the same time. We conclude our analysis by summarising the most important results and providing an outlook.

\section{Lepton Flavoured Scalar Dark Matter}
\label{sec::theory}
In this section we first provide a brief general introduction to the DMFV framework  and then present an explicit simplified model of lepton-flavoured complex scalar DM within DMFV.

\subsection{Dark Minimal Flavour Violation}

In DMFV the SM is extended by a DM flavour triplet $\phi = (\phi_1, \phi_2, \phi_3)^T$ and a corresponding mediator $\psi$. The dark particles are coupled to SM matter through the generic interaction term
\begin{equation}
\lambda_{ij} \bar{f}_i\, \psi\, \phi_j\,, 
\end{equation} 
which is governed by the general $3 \times 3$ complex coupling matrix $\lambda$. Here, $f$ can be either of the SM fermions $u_R, d_R, \ell_R, Q_L$ or $L_L$ and the quantum numbers of $\phi$ and $\psi$ depend on their particle natures and the choice for $f$. This field extension of the SM   extends its approximate global flavour symmetry to
\begin{equation}
\mathcal{G}_\text{DMFV} = U(3)_Q \times U(3)_L \times U(3)_u \times U(3)_d \times U(3)_\ell \times \mathcal{G}(3)_\phi\,,
\end{equation} 
where $\mathcal{G}(3)_\phi$ generally depends on the particle nature of $\phi$. If for instance $\phi$ is a Dirac fermion or a complex scalar $\mathcal{G}(3)_\phi$ is a $U(3)_\phi$, while one is left with an $O(3)_\phi$ symmetry if $\phi$ is chosen to be a Majorana fermion or a real scalar. Together with this extended flavour symmetry, the DMFV ansatz postulates that the coupling matrix $\lambda$ constitutes the only new source of flavour and CP violation beyond the Yukawa couplings $Y_u$, $Y_d$ and $Y_\ell$. The latter assumption is referred to as the DMFV hypothesis.

\subsection{Lepton-Flavoured Scalar Dark Matter in DMFV}
DMFV models with Dirac fermionic DM coupling to right-handed down-type quarks $d_R$ \cite{Agrawal:2014aoa}, right-handed charged leptons $\ell_R$ \cite{Chen:2015jkt}, right-handed up-type quarks $u_R$ \cite{Blanke:2017tnb,Jubb:2017rhm} and the left-handed quark doublets $Q_L$ \cite{Blanke:2017fum}, as well as Majorana fermionic DM coupling to right-handed up-type quarks $u_R$ \cite{Acaroglu:2021qae} have already been studied. For the present analysis we set up a model in the DMFV framework, where DM couples to right-handed charged leptons, i.e.\@ we chose $f = \ell_R$. The DM field $\phi$ is chosen to be a complex scalar, which transforms as a singlet under the SM gauge group with representation $(\mathbf{1},\mathbf{1}, 0)_0$, while the mediator $\psi$ is an electrically charged Dirac fermion with the representation $(\mathbf{1},\mathbf{1},-1)_{1/2}$. Here we have used the short-hand notation $(SU(3)_C,SU(2)_L,U(1)_Y)_\text{spin}$. The Lagrangian of this model reads
\begin{align}
\nonumber
\mathcal{L} = &\mathcal{L}_\text{SM} + (\partial_\mu \phi)^\dagger (\partial^\mu \phi)-M^2_\phi\, \phi^\dagger \phi+ \bar{\psi} (i\slashed{D}-m_\psi)\psi  -(\lambda_{ij} \bar{\ell}_{Ri} \psi\, \phi_j + \text{h.c.})\\
&+ \lambda_{H\phi}\, \phi^\dagger \phi\, H^\dagger H + \lambda_{\phi\phi} \left(\phi^\dagger \phi\right)^2\,.
\label{eq::lagrangian}
\end{align}

As mentioned above, $\lambda$ is a complex $3\times 3$ matrix that couples the DM triplet to the right-handed charged leptons of the SM. The DM field $\phi$ comes in three generations and transforms under a global $U(3)_\phi$ symmetry, as we have chosen it to be a complex scalar. The lightest generation of $\phi$ is assumed to account for the DM abundance in the universe. The mass matrix $M^2_\phi$ as well as the couplings $\lambda_{H\phi}$ and $\lambda_{\phi\phi}$ cannot be general $3 \times 3$ matrices as this would constitute additional sources of flavour violation, which is forbidden by the DMFV hypothesis. Thus, we choose the latter couplings to be diagonal and flavour-universal\footnote{Note that in principle the DMFV ansatz also allows these couplings to be non-diagonal and flavour-violating, if they are parameterised in terms of $\lambda$ following the usual spurion ansatz \cite{DAmbrosio:2002vsn}.}. The DM mass matrix $M^2_\phi$ is discussed in more detail in Section \ref{sec::masscorrections}.\par
Since we choose the DM to be a scalar, the quartic couplings are an additional qualitative feature in our model relative to previous studies of DMFV. The DM-Higgs coupling can mediate tree-level direct detection processes as well as annihilations, and when this coupling is sizeable the phenomenology matches on to the well-studied case of Higgs portal dark matter \cite{Cline:2013gha}. In this paper we will take this coupling to be negligible and focus on the phenomenology of the DM coupling to SM fermions. It will be interesting to study the case when both couplings play an important role, which is left for future work. The self-quartic for DM may in principle induce self-interactions. However, for the parameter space we will find to be viable, a perturbative contact interaction does not have significant self-interactions to affect phenomenology. Therefore, we will ignore both quartic couplings in this work. 
\par
For the parameterization of the coupling matrix $\lambda$, we follow \cite{Agrawal:2014aoa} and perform a singular value decomposition  to find
\begin{equation}
\label{eq::svd}
\lambda = U D V\,,
\end{equation}
where $U$ and $V$ are unitary matrices and $D$ is a diagonal matrix with positive real entries $D_i$. We further use that eq.\@ \eqref{eq::svd} is invariant under a diagonal rephasing of $U$ and $V$ to remove 3 complex phases from $U$. Finally, we remove the unitary matrix $V$ using the flavour symmetry $\mathcal{G}_\text{DMFV}$ and $U(3)_\phi$ in particular, to find
\begin{equation}
\lambda = U D\,. 
\end{equation}
The parametrization of $U$ is adopted from \cite{Blanke:2006xr,Agrawal:2014aoa} and reads
\begingroup
\setlength\arraycolsep{3pt}
	\begin{eqnarray}
	U &=& U_{23}\, U_{13}\, U_{12}\, \nonumber\\
	&=&
 	\begin{pmatrix}
  	1 & 0 & 0\\
 	 0 & c_{23}^\theta & s_{23}^\theta e^{- i\delta_{23}}\\
 	 0 & -s_{23}^\theta e^{i\delta_{23}} & c_{23}^\theta\\
	\end{pmatrix}
	\begin{pmatrix}
 	 c_{13}^\theta & 0 & s_{13}^\theta e^{- i\delta_{13}}\\
 	 0 & 1 & 0\\
 	 -s_{13}^\theta e^{ i\delta_{13}} & 0 & c_{13}^\theta\\
	\end{pmatrix}
	\begin{pmatrix}
 	 c_{12}^\theta & s_{12}^\theta e^{- i\delta_{12}} & 0\\
 	 -s_{12}^\theta e^{i\delta_{12}} & c_{12}^\theta & 0\\
 	 0 & 0 & 1\\
	\end{pmatrix},
	\end{eqnarray}
\endgroup
with the shorthand notation $s^\theta_{ij} = \sin\theta_{ij}$ and $c^\theta_{ij} = \cos\theta_{ij}$.\par
The above parameterization of $\lambda$  contains nine physical parameters
\begin{equation}
\theta_{23},\,\theta_{13},\,\theta_{12},\,\delta_{23},\,\delta_{13},\,\delta_{12},\,D_1,\,D_2,\,D_3\,.
\end{equation}
To avoid a double-counting of the parameter space and to ensure perturbativity we will use the following ranges in our numerical analysis:	
\begin{equation}
\theta_{ij} \in [0,\frac{\pi}{4}], \quad \delta_{ij} \in [0,2\pi), \quad D_i \in [0,2]\,.
\end{equation}
	
\subsection{Mass Spectrum and DM Stability}
\label{sec::masscorrections}
The DMFV hypothesis ensures that there are no DMFV-violating contributions to the mass matrix $M^2_\phi$. However, the UV completion of the model could still induce a DMFV-preserving term at tree or loop level that contributes to the mass splitting between the different DM generations $\phi_i$. Further, there are  one-loop renormalisation contributions to $M^2_\phi$ already within the simplified model. To parameterize these contributions we follow the MFV spurion approach \cite{DAmbrosio:2002vsn} and expand $M^2_\phi$ in powers of $\lambda$ to find
\begin{equation}
\label{eq::masssplitting}
M^2_{\phi,ij} = m_\phi^2 \left\{\mathbbm{1}+ \eta \left(\lambda^\dagger \lambda\right) + \mathcal{O}\left(\lambda^4\right)\right\}_{ij} = m_\phi^2 \left\{\mathbbm{1}+ \eta\, D_i^2 + \mathcal{O}\left(\lambda^4\right)\right\} \delta_{ij}\,.
\end{equation} 
Here we have introduced the parameter $\eta$, which accounts for our ignorance about the details of the UV-complete model. 
\par 
For convenience, in addition we always order the mass eigenstates of the field $\phi$ in such a way that we have
\begin{equation}
M^2_\phi = \text{diag}\left(m_{\phi_1}^2,m_{\phi_2}^2,m_{\phi_3}^2\right)\,,
\end{equation}
with the hierarchy $m_{\phi_1}>m_{\phi_2}>m_{\phi_3}$. This ensures that the third dark generation is always the lightest state.  As we assume it to form the observed DM of the universe, we have to ensure its stability. In Dirac fermionic quark-flavoured DMFV models DM stability is guaranteed by an unbroken residual $\mathbbm{Z}_3$ symmetry which is implied by the global flavour symmetry \cite{Agrawal:2014aoa,Blanke:2017tnb,Blanke:2017fum}. Such a residual symmetry is neither present for lepton-flavoured nor Majorana fermionic DMFV models \cite{Chen:2015jkt, Acaroglu:2021qae}. We thus follow the usual paradigm of imposing a $\mathbbm{Z}_2$ symmetry under which only the new fields $\phi$ and $\psi$ are charged. This ensures that they cannot decay into SM particles only and renders $\phi_3$ stable, as long as its mass $m_{\phi_3}$ is smaller than the charged mediator's mass $m_\psi$. The latter can be guaranteed by choosing the mass corrections to be negative, i.e.\@ $\eta < 0$, and further demanding that
\begin{equation}
m_\phi \leq m_\psi\,,
\end{equation} 
where $m_\phi$ is the tree-level mass parameter from eq.\@ \eqref{eq::masssplitting}.

\section{Collider Phenomenology}
\label{sec::collider}
In this section we analyse the constraints that current collider searches place on the model presented above. In doing so, we  mainly discuss LHC limits, while limits that the LEP experiments place on our model are taken into account by always choosing $m_\psi > 100\, \mathrm{GeV}$ \cite{ALEPH:2002gap,DELPHI:2003uqw}. 

\subsection{Relevant LHC Signatures}
\label{sec::lhcsignatures}
\begin{figure}[t!]
	\centering
		\begin{subfigure}[t]{0.4\textwidth}
		\includegraphics[width=\textwidth]{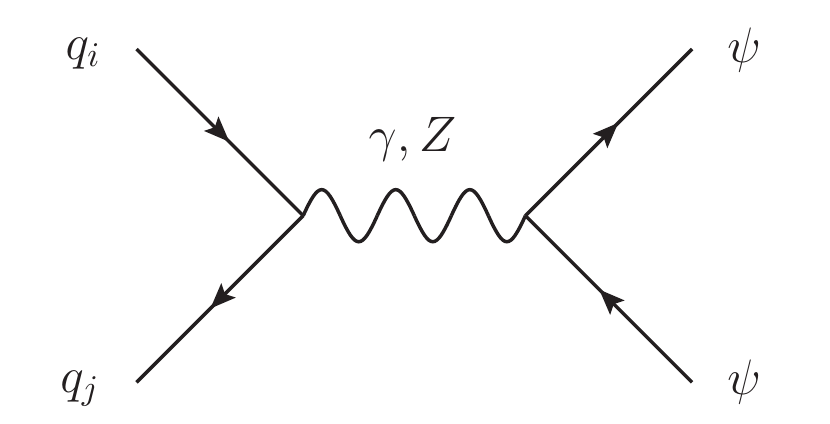}
		\caption{$\psi\bar{\psi}$ production}
		\label{fig::psiproduction}
		\end{subfigure}
		\hspace*{1cm}
		\begin{subfigure}[t]{0.4\textwidth}
		\includegraphics[width=\textwidth]{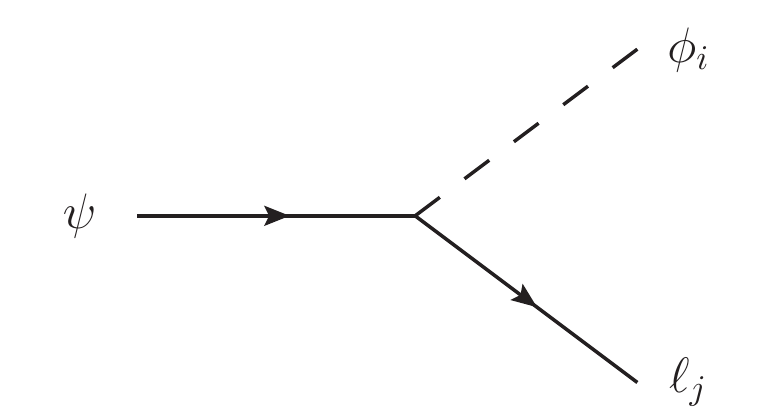}
		\caption{$\psi$ decay}
		\label{fig::psidecay}
		\end{subfigure}
	\caption{Feynman diagrams of $\psi$ pair-production and its subsequent decay.}
	\end{figure}

Pair-production of the mediator $\psi$ takes place via a Drell-Yan process, where a quark and an anti-quark from the initial-state protons annihilate into an off-shell $\gamma$ or a $Z$ that decays into $\psi\bar{\psi}$, as shown in Figure \ref{fig::psiproduction}. The mediator's subsequent decay into a lepton $\ell$ and a dark scalar $\phi$ (see Figure \ref{fig::psidecay}) will then result in the final state $\ell_i \bar{\ell}_j \phi^\dagger_k\phi_l$, where $i,j,k$ and $l$ are flavour indices. Note that the $\mathbbm{Z}_2$ symmetry introduced in Section \ref{sec::masscorrections} ensures that $\psi$ always decays into final states that include  $\phi$. Putting things together, we find the following process to be relevant at the LHC:
\begin{equation}
pp \,\,\, \rightarrow \,\,\, \psi \bar{\psi} \,\,\, \rightarrow \,\,\, \ell_i \bar{\ell}_j \phi^\dagger_k \phi_l \,.
\end{equation}
This process gives rise to the same-flavour signatures $\ell_i \bar{\ell}_i + \slashed{E}_T$ as well as the mixed-flavour signatures $\ell_i \bar{\ell_j} + \slashed{E}_T$.\footnote{Note that due to the small mass-splitting between the different dark flavours, we can safely neglect the soft visible decay products of the heavier states and assume that all three states $\phi_k$ appear as missing transverse energy.} The latter signatures are typically neglected in LHC searches, since in non-flavoured models of DM they are correlated with the strongly constrained lepton flavour violating (LFV) decays $\ell_i \rightarrow \ell_j \gamma$. In flavoured DM models however, the mixed-flavour signatures are proportional to the diagonal elements of the coupling matrix governing the interaction between DM and the SM. In such models, we can get the mixed $\ell_i \ell_j +\slashed{E}_T$  at a sizeable level without flavor violation (unlike e.g.~supersymmetric models with neutralino dark matter) since dark particles can carry away the flavor quantum number. This is similar to the SM $W^+W^-$ production where neutrinos carry away the flavor quantum numbers. We will return to these signatures when performing a combined analysis in Section \ref{sec::combined}.
\par
To constrain our model we  use the experimental limits on the final states $e\bar{e}  + \slashed{E}_T$, $\mu\bar{\mu}  + \slashed{E}_T$ and $\tau\bar{\tau}  + \slashed{E}_T$, where the first two are constrained by searches for supersymmetric scalar leptons (sleptons) of the first and second generations and the latter is contrained by searches for stau leptons. Note that the final-state kinematics could in principle be different due to distinct spin-statistics -- while in SUSY models the sleptons $\tilde{\ell}$ are scalar particles that decay into two fermions, in our model the fermionic mediator $\psi$ decays into another fermion $\ell$ and scalar DM $\phi$. We do not expect efficiencies of the analysis to be strongly kinematic dependent, and assume that they stay the same for our case. Since the overall cross section for the fermion pair-production is higher than that for scalars in Drell-Yan, we will implement the production explicitly and compare our cross section with the limits on SUSY models. In what follows we use the short-hand notation $\ell\bar{\ell}+\slashed{E}_T$ for the joint signatures $e\bar{e}  + \slashed{E}_T$ and $\mu\bar{\mu}  + \slashed{E}_T$.

\subsection{Recast of LHC Limits}
\label{sec::lhclimits}
Experimental searches for the $\ell\bar{\ell}+\slashed{E}_T$ signature have been carried out at $13\,\mathrm{TeV}$~\cite{CMS:2020bfa,ATLAS:2019lff, CMS:2018eqb} and also $8\,\mathrm{TeV}$ \cite{ATLAS:2014zve}, where the most stringent constraints are placed by the CMS search \cite{CMS:2020bfa} based on the full LHC run 2 data set with an integrated luminosity of $137\,\mathrm{fb}^{-1}$. For the signature $\tau\bar{\tau} + \slashed{E}_T$ we use the ATLAS search \cite{ATLAS:2019gti} that again is based on the full run 2 data set with an integrated luminosity of $139\,\mathrm{fb}^{-1}$.\par 
To recast these searches we implement our model in \texttt{FeynRules} \cite{Alloul:2013bka} using the Lagrangian of eq.\@ \eqref{eq::lagrangian}. We then use this implementation to generate a \texttt{MadGraph} \cite{Alwall:2014hca} model and calculate the leading-order (LO) signal cross section $\sigma \times \mathrm{Br}$ for the signatures $\ell \ell + \slashed{E}_T$ and $\tau \tau + \slashed{E}_T$ separately to compare it with the respective upper limits obtained from the searches mentioned above.  Note that we follow \cite{Agrawal:2014aoa,Acaroglu:2021qae, Blanke:2017tnb,Blanke:2017fum} and neglect the mass splitting between the different DM flavours $\phi_i$ described in Section \ref{sec::masscorrections} and the resulting decay of the heavier flavours into the lightest state, as the smallness of the splitting results in soft and therefore hard to detect decay products. As in our earlier work we also set the mixing angles and phases in $\lambda$ to zero, as we are primarily interested in the constraints that the LHC searches place on the $m_\psi - m_\phi$ plane. Since non-vanishing mixing angles reduce the branching ratio of a given flavour-conserving final state and thus reduce its signal cross section, setting the mixing angles and phases to zero only further strengthens the constraints placed on the mass plane. We also adopt the choice of setting $D_1 = D_2$ when analyzing the LHC constraints, in order to straightforwardly recast the limits from the searches mentioned above.\par
\begin{figure}[b!]
	\centering
		\begin{subfigure}[t]{0.49\textwidth}
		\includegraphics[width=\textwidth]{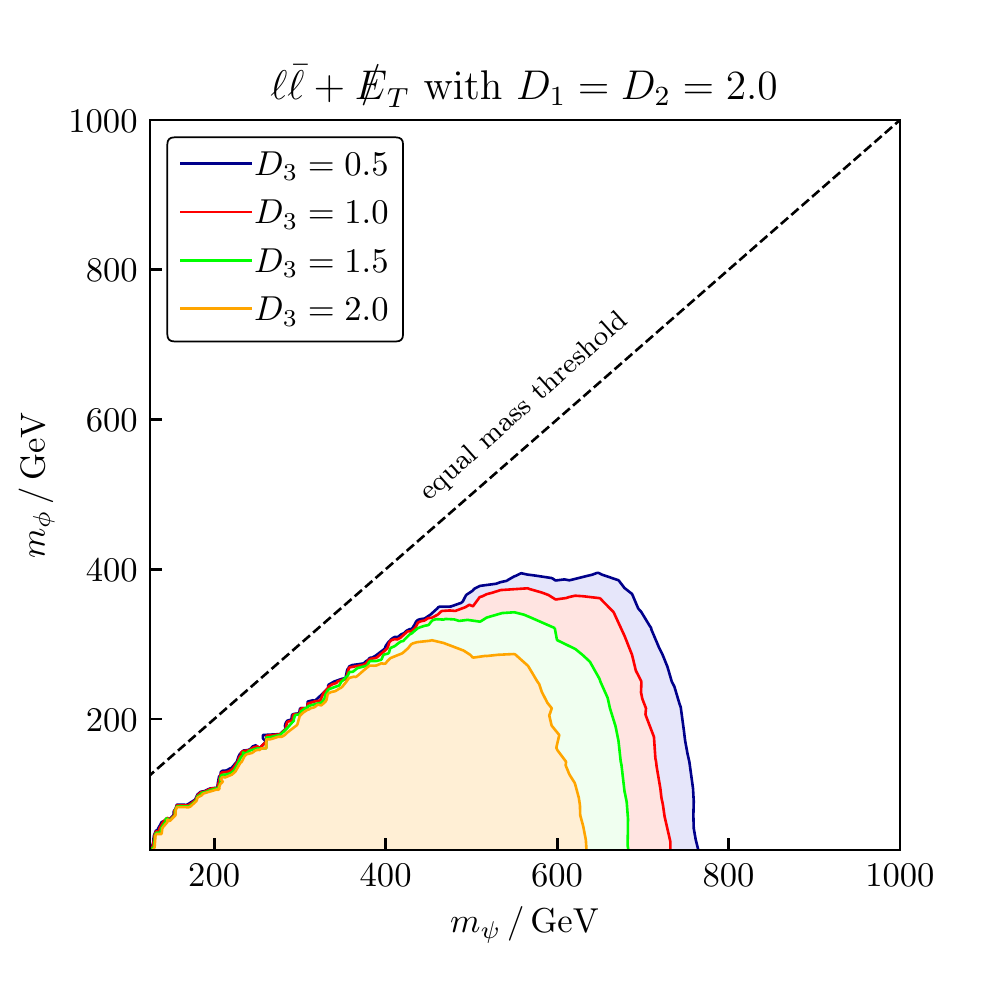}
		\caption{final states with $\ell\bar{\ell} + \slashed{E}_T$}
		\label{fig::dilep}
		\end{subfigure}
		\hfill
		\begin{subfigure}[t]{0.49\textwidth}
		\includegraphics[width=\textwidth]{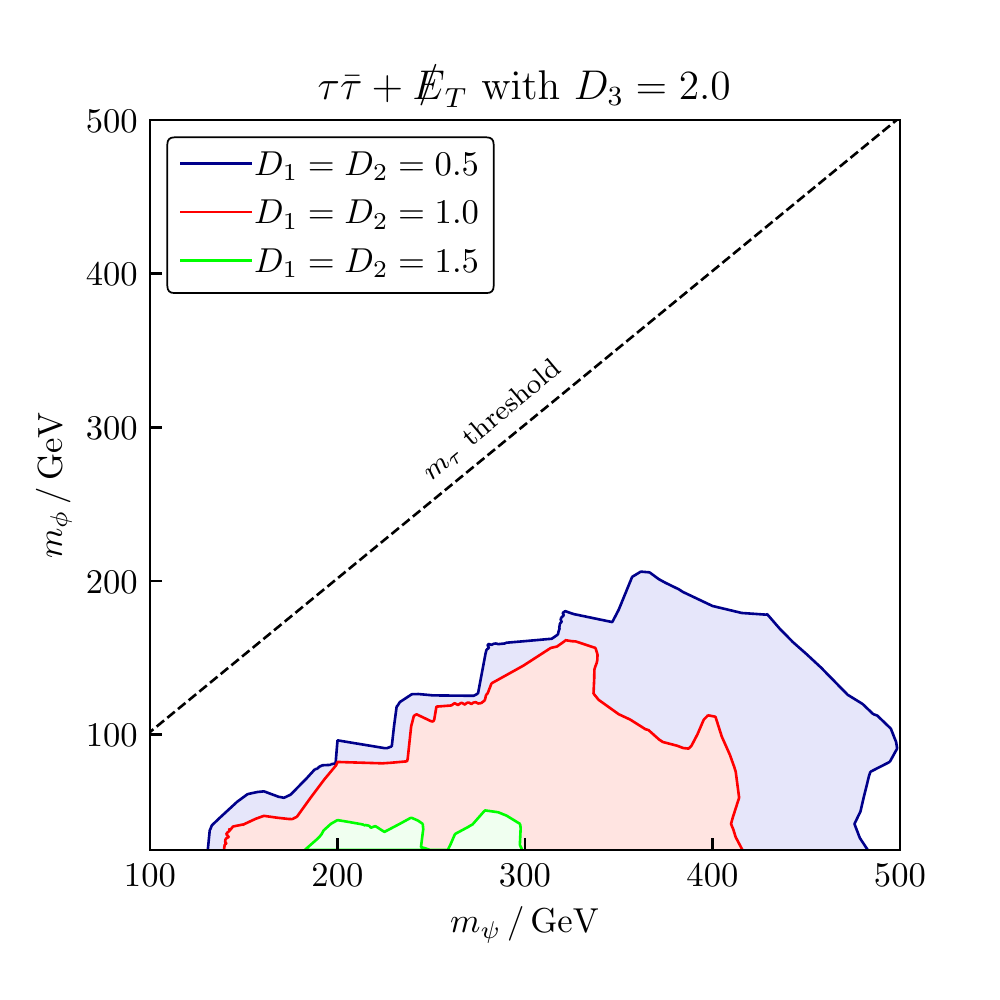}
		\caption{final states with $\tau\bar{\tau} + \slashed{E}_T$}
		\label{fig::tau}
		\end{subfigure}
	\caption{Constraints on the final states $\ell\bar{\ell} + \slashed{E}_T$ and $\tau\bar{\tau} + \slashed{E}_T$. The areas under the curves are excluded.}
	\label{fig::lhcconstraints}
	\end{figure}

The results of this recasting procedure are shown in Figure \ref{fig::lhcconstraints}. In Figure \ref{fig::dilep} we show the constraints that final states with $\ell\bar{\ell} + \slashed{E}_T$ place on the $m_\psi - m_\phi$ plane. The largest area is excluded for the smallest value of $D_3$. As growing values of $D_3$ increase the branching ratio of the process $\psi \rightarrow \tau\phi^\dagger$, they also decrease the branching ratio of the relevant process $\psi \rightarrow \ell \phi^\dagger$ which in turn leads to a smaller signal cross section. 
  Thus, the excluded area shrinks with a growing coupling $D_3$. For maximal couplings to electrons and muons $D_1 = D_2 = 2.0$ and a small coupling to the tau $D_3 = 0.5$ (blue line), the constraints are thus most stringent and enforce either $m_\psi \GtrSim 750\,\mathrm{GeV}$ while $m_\phi$ can be chosen freely or $m_\psi \GtrSim 400\,\mathrm{GeV}$ and $m_\phi \GtrSim 400\,\mathrm{GeV}$.\par 
  In Figure \ref{fig::tau} one can see that searches for the final state $\tau\bar{\tau} + \slashed{E}_T$ generally place much weaker constraints on the mass plane. Again and for the same reason as above, the excluded area is the largest when the coupling to electrons and muons $D_1 = D_2$ is small compared to $D_3$. However, since in this case raising $D_1 = D_2$ results in the growth of the branching ratio of both processes $\psi \rightarrow e\phi^\dagger$ and $\psi \rightarrow \mu\phi^\dagger$, the signal cross section drops much more quickly and the excluded area shrinks much faster, such that for the near-degenerate case of $D_1 = D_2 \approx D_3$ there is close to no exclusion (green line). For the case of a maximal exclusion (blue line) the constraints can be fulfilled by choosing either $m_\psi \GtrSim 500\,\mathrm{GeV}$ and a free $m_\phi$ or $m_\psi \GtrSim 200\,\mathrm{GeV}$ and $m_\phi \GtrSim 200\,\mathrm{GeV}$.

\section{Flavour Physics Phenomenology}
\label{sec::flavour}

In DMFV, the unrestricted structure of the flavour violating coupling matrix $\lambda$ can generally lead to large flavour changing neutral currents (FCNC). As in our model the DM triplet $\phi$ couples to leptons, they give rise to sizeable contributions to the aforementioned LFV decays $\ell_i\to\ell_j\gamma$. Identifying flavour-safe scenarios thus puts strong constraints on the parameter space of our model. We  use this section to determine such scenarios and find the corresponding viable parameter space of our model.

\subsection{Lepton Flavour Violating Decays}
\begin{figure}[t!]
	\centering
	\includegraphics[width=0.5\textwidth]{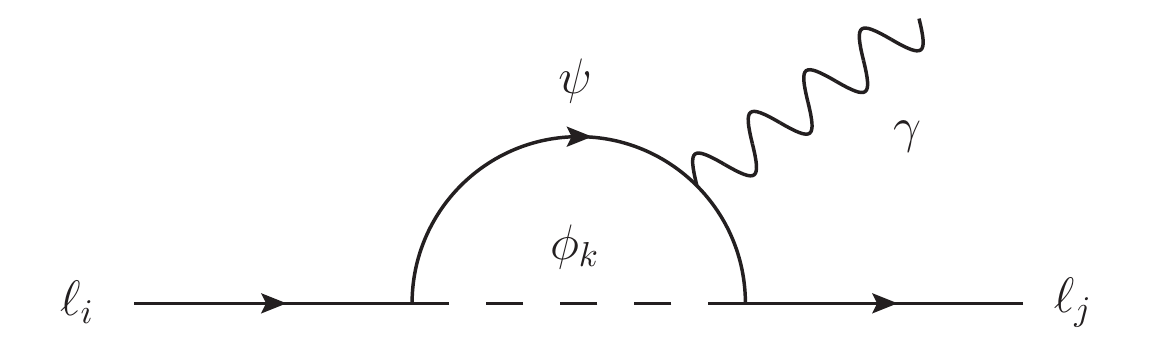}
	\caption{Feynman diagram for the LFV decay $\ell_i \rightarrow \ell_j \gamma$. The contribution from the photon coupling to one of the SM leptons is not shown.}
	\label{fig::lfv}
\end{figure}

The LFV decay $\ell_i \rightarrow \ell_j \gamma$ shown in Figure \ref{fig::lfv} is governed by the interaction term of the Lagrangian in eq.\@ \eqref{eq::lagrangian} and is proportional to the off-diagonal elements of the coupling matrix $\lambda$. Following \cite{Kersten:2014xaa} we express its amplitude as
\begin{equation}
\mathcal{M}_{\ell_i \ell_j \gamma} = \frac{e}{2 m_{\ell_i}} \epsilon^{*\alpha}\bar{u}_{\ell_j}\left[i\sigma_{\beta\alpha}q^\beta\left(a_{\ell_i \ell_j \gamma}^R P_L + a_{\ell_i \ell_j \gamma}^L P_R\right) \right]u_{\ell_i}\,,
\end{equation}
where $\sigma_{\beta\alpha} = i[\gamma_\alpha,\gamma_\beta]/2$, $\epsilon^\alpha$ is the photon polarization vector and $P_{R/L} = (1\pm\gamma_5)/2$ are projection operators. Note that we adopt the convention \cite{Hisano:1995cp, Chacko:2001xd} where the superscript of the coefficients $a^{R/L}_{\ell_i \ell_j \gamma}$ refers to the chirality of the final state. For a generic interaction of the form 
\begin{equation}
\label{eq::genericint}
\mathcal{L}_\text{int} = c^R_{ij}\, \bar{\ell}_{Ri} \psi \phi_j + c^L_{ij}\, \bar{\ell}_{Li} \psi \phi_j + \text{h.c.}\,,
\end{equation}
with mass parameters $m_\psi$ and $m_{\phi_i}$ their expressions generally read \cite{Chacko:2001xd,Kersten:2014xaa}
\begin{align}
\label{eq::lfvaRgeneral}
a_{\ell_i \ell_j \gamma}^R &= \frac{m_{\ell_i}}{16\pi^2}\sum_k\left(\frac{m_{\ell_i}}{12 m_{\phi_k}^2 }c^{R*}_{ik} c^R_{jk} F(x_k) + \frac{m_\psi}{3m_{\phi_k}^2} c^{L*}_{ik} c^R_{jk} G(x_k)\right)\,,\\
\label{eq::lfvaLgeneral}
a_{\ell_i \ell_j \gamma}^L &= \frac{m_{\ell_i}}{16\pi^2}\sum_k\left(\frac{m_{\ell_i}}{12 m_{\phi_k}^2 }c^{L*}_{ik}c^L_{jk}  F(x_k) + \frac{m_\psi}{3m_{\phi_k}^2} c^{R*}_{ik} c^L_{jk} G(x_k)\right)\,, 
\end{align}
with $x_k = m_\psi^2/m_{\phi_k}^2$. Here, $\psi$ is a Dirac fermion with electric charge $Q_\psi = -1$ and the $\phi_i$ are scalars. The loop functions $F$ and $G$ are given as \cite{Chacko:2001xd,Kersten:2014xaa}
\begin{align}
\label{eq::lfvloopfcts}
F(x) &= \frac{2}{(1-x)^4}\left[2+3 x-6 x^2+x^3+6 x \ln x\right]\,,\\
G(x) &= -\frac{3}{2 (1-x)^3} \left[3-4x+x^2+2 \ln x\right]\,.
\end{align}
In terms of these coefficients, the branching ratio of LFV decays is given by
\begin{equation}
\label{eq::lfvBRgeneral}
\text{BR}(\ell_i \rightarrow \ell_j \gamma) = \frac{e^2}{64\pi} \frac{m_{\ell_i}}{\Gamma_{\ell_i}} \left(|a_{\ell_i \ell_j \gamma}^R|^2+|a_{\ell_i \ell_j \gamma}^L|^2\right)\,,
\end{equation}
where $\Gamma_{\ell_i}$ denotes the total decay width of $\ell_i$.\par 
Since in our model the DM flavour triplet $\phi$ only couples to right-handed leptons, the expression from above reduces to
\begin{equation}
\label{eq::lfvBR}
\text{BR}(\ell_i \rightarrow \ell_j \gamma) = \frac{e^2}{64\pi} \frac{m_{\ell_i}}{\Gamma_{\ell_i}} |a_{\ell_i \ell_j \gamma}^R|^2\,,
\end{equation}
and the contributions to $a_{\ell_i \ell_j \gamma}^R$ from diagrams with a chirality flip inside the loop vanish, i.e.\@
\begin{equation}
\label{eq::lfvaR}
a_{\ell_i \ell_j \gamma}^R = \frac{m_{\ell_i}^2}{192\pi^2}\sum_k\frac{\lambda^*_{ik} \lambda_{jk}}{m_{\phi_k}^2 } F(x_k)\,.
\end{equation}

\subsection{Constraints from LFV Decays}

In order to apply the limits from LFV decays to our model we will generate random points in the parameter space, calculate the branching ratios from eq.\@ \eqref{eq::lfvBR} and demand that they are smaller than their experimental upper limits. The latter are given at $90\%$ C.L.\@ and read \cite{MEG:2016leq,BaBar:2009hkt,Belle:2021ysv}
\begin{alignat}{2}
&\text{BR}(\mu \rightarrow e \gamma)_\text{max} &&= 4.2 \times 10^{-13}\,,\\
&\text{BR}(\tau \rightarrow e \gamma)_\text{max} &&= 3.3 \times 10^{-8\phantom{1}}\,,\\
&\text{BR}(\tau \rightarrow \mu \gamma)_\text{max} &&= 4.2 \times 10^{-8\phantom{1}}\,.
\end{alignat}
For our numerical analysis, the values for lepton masses $m_{\ell_i}$ and decay widths $\Gamma_{\ell_i}$ are obtained from \cite{ParticleDataGroup:2020ssz}. We again follow \cite{Agrawal:2014aoa,Blanke:2017fum,Acaroglu:2021qae, Blanke:2017tnb} and neglect the mass corrections discussed in Section \ref{sec::masscorrections}, as they would lead to higher-order DMFV corrections that we assume to be negligible. In this limit, eq.\@ \eqref{eq::lfvaR} reduces to
\begin{equation}
a_{\ell_i \ell_j \gamma}^R = \frac{m_{\ell_i}^2}{192\pi^2}\frac{\left(\lambda\lambda^\dagger\right)_{ji}}{m_\phi^2 } F(x)\,.
\end{equation}\par
Using the experimental upper limits quoted above, one can already estimate how strong the constraints on the off-diagonal elements of $\lambda\lambda^\dagger$ are. For the most stringently constrained LFV decay $\mu \rightarrow e \gamma$, for example, we find
\begin{equation}
\label{eq::lfvest}
\sqrt{|\left(\lambda\lambda^\dagger\right)_{\mu e}|} \LessSim \frac{m_\psi}{15 \,\mathrm{TeV}}\,,
\end{equation}
where we have expanded eq.\@ \eqref{eq::lfvBR} for $m_\psi \gg m_{\phi}$. Thus, for mediator masses $m_\psi$ of order $\mathcal{O}(1\, \mathrm{TeV})$ the coupling matrix $\lambda$ has to satisfy 
\begin{equation}
\sqrt{|\left(\lambda\lambda^\dagger\right)_{\mu e}|} \sim \mathcal{O}(0.01-0.1)\,.
\end{equation}\par
\begin{figure}[b!]
	\centering
		\begin{subfigure}[t]{0.49\textwidth}
		\includegraphics[width=\textwidth]{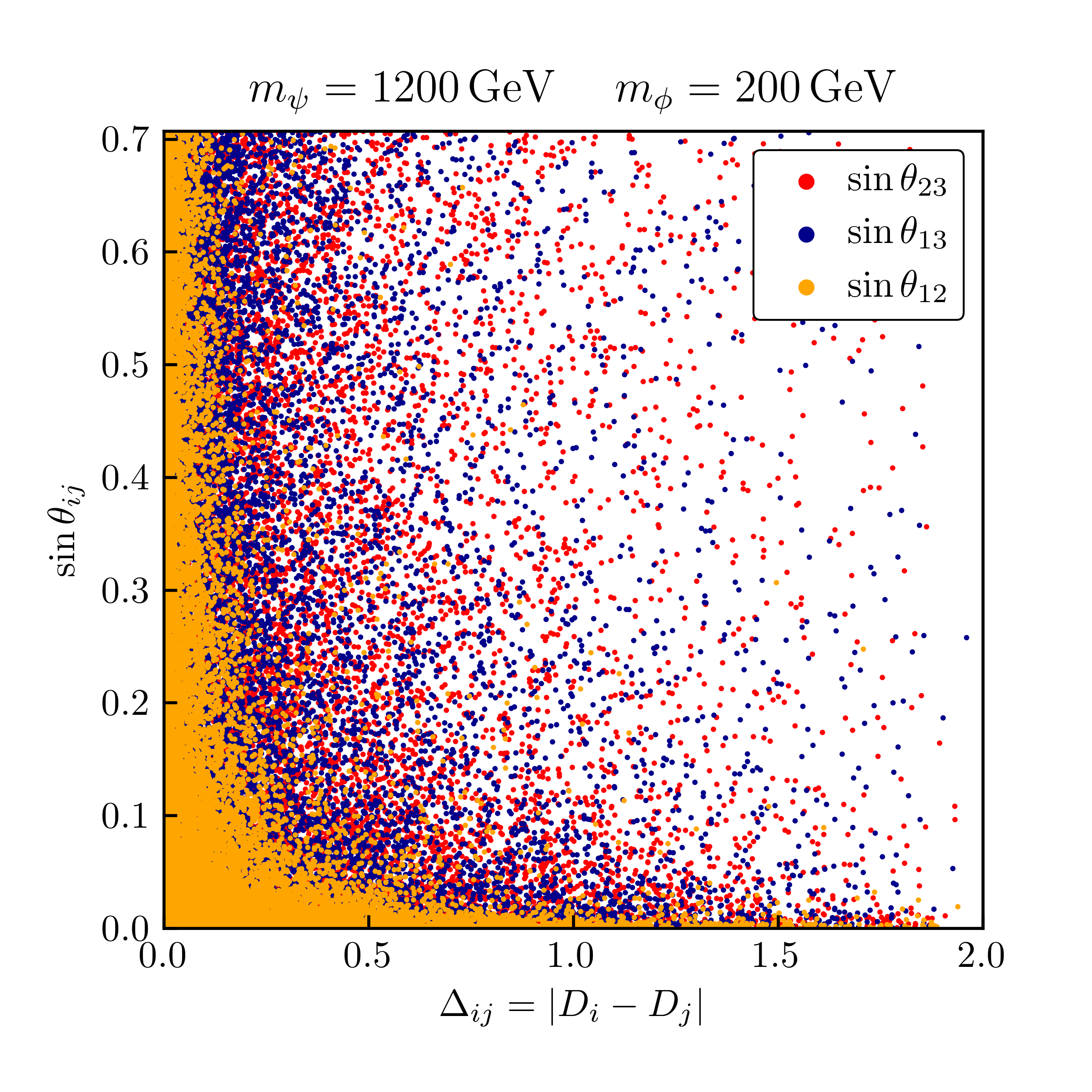}
		\end{subfigure}
		\hfill
		\begin{subfigure}[t]{0.49\textwidth}
		\includegraphics[width=\textwidth]{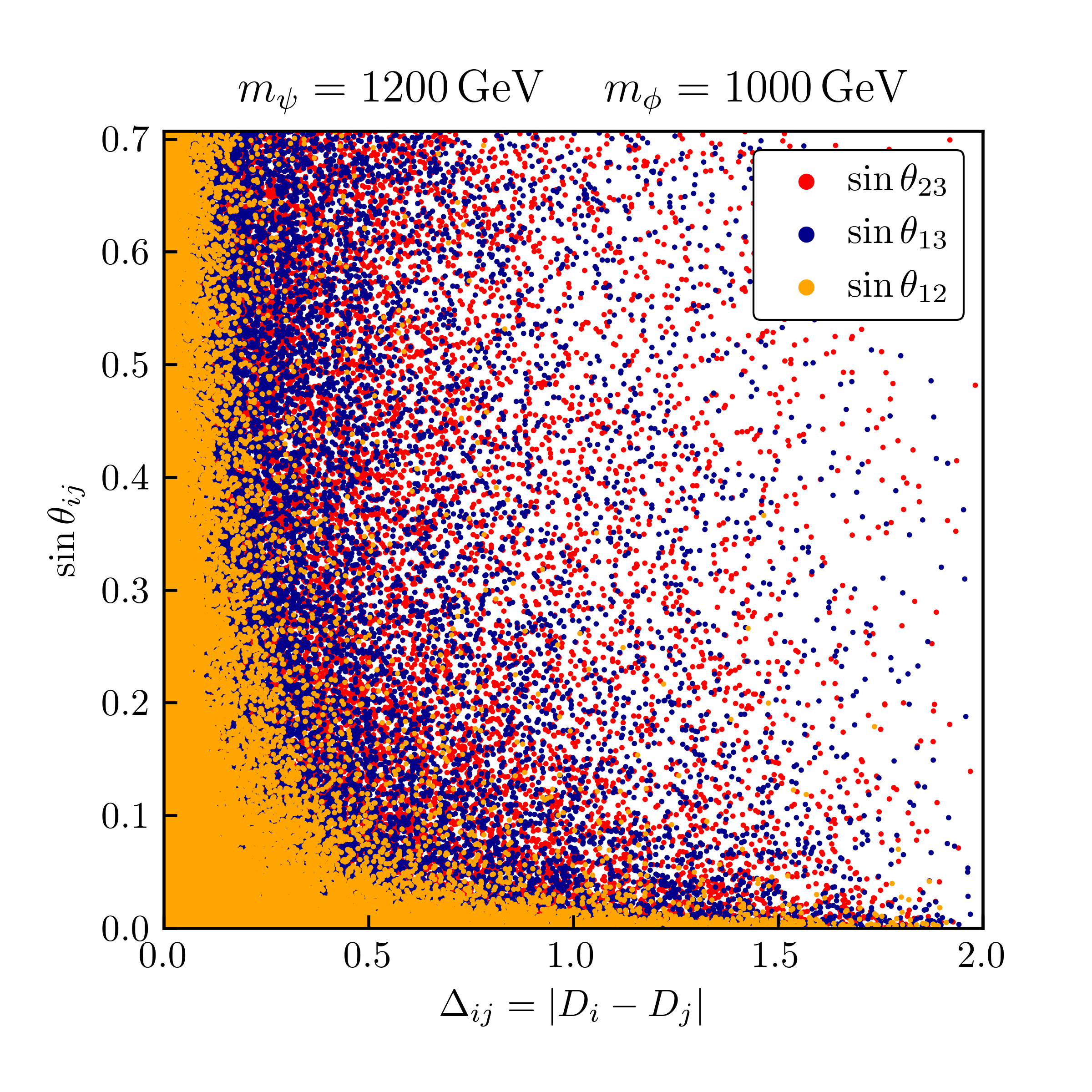}
		\end{subfigure}
	\caption{Allowed mixing angles $\theta_{ij}$ in dependence of the splittings $\Delta_{ij} = |D_i - D_j|$ for $m_\psi = 1200\,\mathrm{GeV}$ and two choices of $m_\phi$.}
	\label{fig::flavourconstraints}
	\end{figure}

In Figure \ref{fig::flavourconstraints} we show the results of our numerical analysis.
We find, as expected, that the LFV decays mainly constrain the mixing angle $\theta_{12}$ while $\theta_{13}$ and $\theta_{23}$ can be chosen freely even for large splittings $\Delta_{ij} = |D_i - D_j|$. Generally, all mixing angles $\theta_{ij}$ exhibit the same dependence on the corresponding splittings $\Delta_{ij}$, as they can take arbitrarily large values for sufficiently small $\Delta_{ij}$. This is due to the fact that the product $\lambda\lambda^\dagger$ becomes diagonal in the degeneracy limit $D_i = D_0$, i.e.\@
\begin{equation}
\lambda\lambda^\dagger = D_0^2 \mathbbm{1}\,.
\end{equation}
Note that in contrast to our estimate of eq.\@ \eqref{eq::lfvest} which we derived for the limit $m_\psi \gg m_\phi$, the LFV constraints actually carry a DM mass dependence. As the DM mass approaches the mediator mass, $a^R_{\mu e \gamma}$ decreases, leading to larger allowed values for $\Delta_{12}$ with a free corresponding mixing angle $\theta_{12}$, as can be seen in Figure \ref{fig::flavourconstraints}. We conclude that the constraints from LFV decays can be satisfied by either choosing a near degeneracy between the couplings to electrons and muons or by choosing arbitrary values for $D_1$ and $D_2$ and suppressing the mixing angle $\theta_{12}$ sufficiently.\par
Before completing this section we want to also comment on precision measurements of leptonic electric dipole moments (EDM) $d_\ell$ as well as magnetic dipole moments (MDM) $a_\ell$ and their implications for the model presented in this study. Generally, both quantities are induced by the radiative process shown in Figure \ref{fig::lfv} with $i=j$. However, due to the chiral coupling structure in eq.\@ \eqref{eq::lagrangian}, i.e.\@ due to the fact that the DM flavour triplet only couples to right-handed leptons, our model does not induce a contribution to $d_\ell$ at the one-loop level. For the same reason it also lacks the possibility of a chirality enhancement of $a_\ell$. This means that sizeable contributions to $a_\ell$ can only be generated for mediator masses $m_\psi$ of order $\mathcal{O}(100 \,\mathrm{GeV})$, which in turn are excluded by the collider searches we have discussed in Section \ref{sec::collider}. Thus, our model is neither capable of explaining the long-standing tension between the experimental measurement of the muon MDM $a_\mu$ \cite{Muong-2:2006rrc,Muong-2:2021ojo} and its theoretical prediction \cite{Aoyama:2020ynm}, nor do the MDMs \cite{PhysRevLett.100.120801,Muong-2:2021ojo,DELPHI:2003nah} or EDMs \cite{ACME:2018yjb,Muong-2:2008ebm,Belle:2002nla} of charged leptons put relevant constraints on it.

\section{DM Relic Density}
\label{sec::relicdensity}
As we assume the particle $\phi_3$ to account for the observed DM in the universe, demanding from our model to yield the correct DM relic density \cite{Planck:2018vyg} again places constraints on its parameter space. This section is dedicated to the analysis of these constraints.

\subsection{DM Thermal Freeze-Out}
To calculate the DM relic density we assume a thermal freeze-out at $m_{\phi_3}/T_f \approx 20$, i.e.\@ at that time the production as well as annihilation rates of DM fall below the Hubble expansion rate, such that the dark species decouples from thermal equilibrium and one is left with a relic amount of DM. The resulting relic number density of DM  depends on its effective annihilation rate $\langle\sigma v\rangle_\text{eff}$.\par 
In our model, the details of the freeze-out in terms of the presence of different dark flavours $\phi_i$ are determined through their mass hierarchy generated by the mass splittings discussed in Section \ref{sec::masscorrections}. Here, we follow \cite{Agrawal:2014aoa,Blanke:2017fum,Blanke:2017tnb,Acaroglu:2021qae} and identify two distinct benchmark scenarios for the freeze-out. The first is the so called \textit{Quasi-Degenerate Freeze-Out} (QDF), where we assume the splittings between the heavier states and the lightest state to be very small. This ensures that the decay of the heavier states into the lightest state is kinematically suppressed to an extent that all dark flavors $\phi_i$ are stable on the freeze-out timescale, and contribute equally to the freeze-out. The heavy DM states will then decay into the lightest state at late times\footnote{Following the line of argument in Appendix D of \cite{Agrawal:2014aoa}, it can be shown that these decays happen at a fast enough rate such that they are not subject to constraints from big bang nucleosynthesis or energy injections into the cosmic microwave background.}. The \textit{Single-Flavour Freeze-Out} (SFF) on the other hand, is defined by a significant mass splitting between $\phi_{1,2}$ and the lightest state $\phi_3$. This in turn leads to a short lifetime of the heavy states compared to the time of the freeze-out. While flavour changing scattering processes between the different DM generations maintain a relative equilibrium amongst them, the number density of the heavier states are suppressed by a Boltzmann factor with the respective mass splitting as its argument. Thus, we assume that in the SFF scenario only the lightest state $\phi_3$ contributes to the freeze-out. Numerically we define these two scenarios as follows:
\begin{itemize}
\item For the QDF scenario we demand that the mass splitting 
\begin{equation}
\Delta m_{i3} = \frac{m_{\phi_i}}{m_{\phi_3}}-1\,
\end{equation}
between the heavier states with $i \in \{1,2\}$ and the lightest state is smaller than $1\%$. To this end we set $\eta = -0.01$ in eq.\@ \eqref{eq::masssplitting}, where smaller values are not reasonable as the splitting may be induced by one-loop diagrams.
\item To render the approximation of the SFF scenario accurate, we demand that the mass splittings $\Delta m_{i3}$ are larger than $10\%$. Note that however the splitting may not grow arbitrarily large in order to ensure convergence of the DMFV expansion, eq.\@ \eqref{eq::masssplitting}. We choose $\eta = - 0.075$ which leads to maximal splittings of $\Delta m_{i3} \simeq 30\%$. 
\end{itemize}

\begin{figure}[b!]
	\centering
		\begin{subfigure}[t]{0.32\textwidth}
		\includegraphics[width=\textwidth]{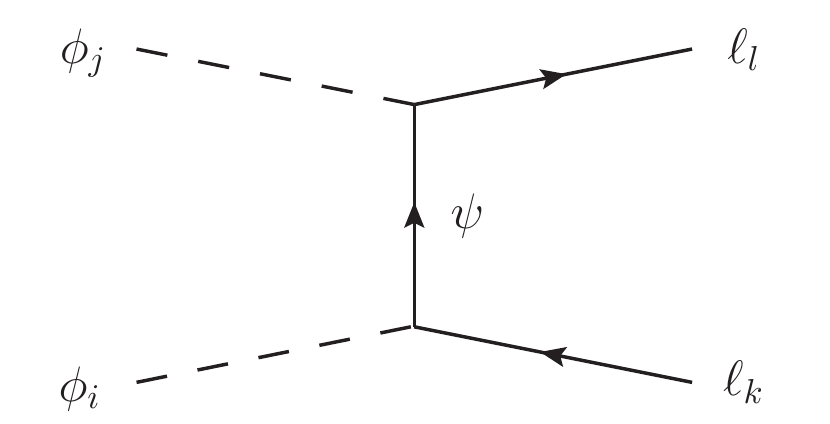}
		\caption{$\phi_i\phi_j$ annihilation}
		\label{fig::annihilation1}
		\end{subfigure}
		\hfill
		\begin{subfigure}[t]{0.32\textwidth}
		\includegraphics[width=\textwidth]{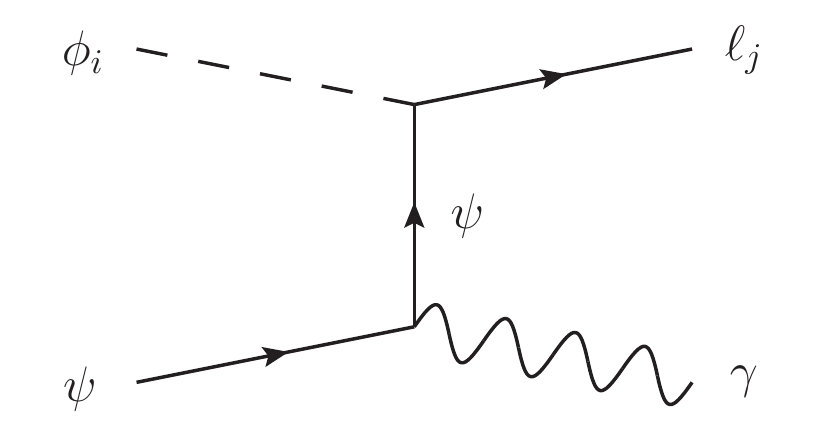}
		\caption{$\phi_i \psi$ coannihilation}
		\label{fig::annihilation2}
		\end{subfigure}
		\hfill
		\begin{subfigure}[t]{0.32\textwidth}
		\includegraphics[width=\textwidth]{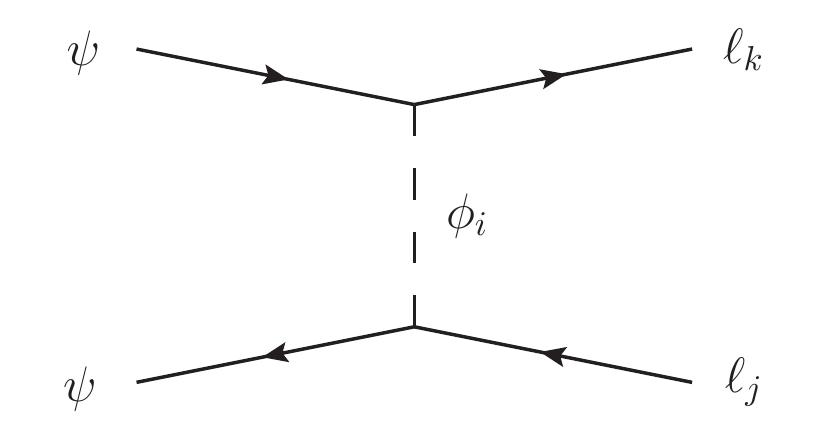}
		\caption{$\psi \psi$ annihilation}
		\label{fig::annihilation3}
		\end{subfigure}
	\caption{Representative Feynman diagrams for annihilations of the new particles into SM matter.}
	\label{fig::annihilation}
	\end{figure}

In Figure \ref{fig::annihilation} we show the tree-level diagrams of the relevant annihilation processes. The coannihilation from Figure \ref{fig::annihilation2} is suppressed by the Boltzmann factor 
\begin{equation}
k = e^{-\frac{m_\psi - m_{\phi_3}}{T_f}}\simeq e^{-20 \frac{m_\psi - m_{\phi_3}}{m_{\phi_3}}}\,,
\end{equation}
and the smallness of the fine structure constant $\alpha_\text{em}$, while the process shown in Figure \ref{fig::annihilation3} receives an even stronger  suppression by a factor $k^2$. Hence the latter processes only become relevant in the nearly degenerate region $m_\psi \approx m_\phi$. Note that the coupling $\lambda_{H\phi}$ from eq.\@ \eqref{eq::lagrangian} in general also leads to relevant annihilations, where two dark particles annihilate into a (virtual) Higgs boson in the $s$-channel, which then decays into SM fermions. The process with two top quarks in the final state can yield a sizeable contribution proportional to $y_t^2 \lambda_{H\phi}^2$ due to the large top Yukawa coupling $y_t$. However, as we are primarily interested in the structure of the flavour-violating coupling $\lambda$ in this analysis and as we do not constrain $\lambda_{H\phi}$ throughout this paper, we use this freedom and assume that the annihilation process via an $s$-channel Higgs boson can be neglected.\par
 
Thus, for the calculation of the annihilation cross section we only consider the $t$-channel annihilation shown in Figure \ref{fig::annihilation1}. For the corresponding diagram we find
\begin{equation}
\overline{|M|^2} = \frac{1}{9}\sum_{ij}\sum_{kl}|\lambda_{ik}|^2|\lambda_{jl}|^2 \frac{\left(m_{\phi_j}^2-m_{\ell_l}^2-t\right) \left(t+m_{\ell_k}^2-m_{\phi_i}^2\right)- t\left(s - m_{\ell_k}^2 - m_{\ell_l}^2\right)}{(t-m_\psi^2)^2}\,,
\end{equation}
for the flavour-averaged squared amplitude $\overline{|M|^2}$. The Mandelstam variables are defined as $s = (p_1 + p_2)^2$ and $t = (p_1 - p_3)^2$, and we sum over the flavour indices $i,j,k$ and $l$. 	
To calculate the effective thermally averaged annihilation cross section $\langle\sigma v\rangle_\text{eff}$ we perform a low-velocity expansion \cite{Srednicki:1988ce, Gondolo:1990dk} yielding 
\begin{align}
\langle\sigma v\rangle_\text{eff} = \frac{1}{2}\langle\sigma v\rangle = \frac{1}{2} \left(a + b \,\langle v^2\rangle + \mathcal{O}(\langle v^4\rangle)\right)\,,
\label{eq::partialwave}
\end{align} 
where $\langle v^2\rangle = 6 T_f/m_{\phi_3} \simeq 0.3$. The factor of two for the conversion between $\langle\sigma v\rangle_\text{eff}$ and $\langle\sigma v\rangle$ is due to $\phi$ being a complex scalar. Using the procedure provided in \cite{Gondolo:1990dk, Wells:1994qy} we calculate the coefficients of the low-velocity expansion. In the limit of equal initial state masses $m_{\phi_i} = m_\phi$ and zero final state masses $m_{\ell_i}=0$ they are found to be
\begin{align}
\label{eq::partiala}
a &= 0\,,\\
\label{eq::partialb}
b &= \frac{1}{9}\sum_{ij}\sum_{kl}\frac{|\lambda_{ik}|^2|\lambda_{jl}|^2}{48 \pi m_\psi^2} \frac{\mu}{(1+\mu)^2}\,,
\end{align} 
with $\mu = m_\phi^2/m_\psi^2$, i.e.\@ we encounter a $p$-wave suppression of $\langle\sigma v\rangle_\text{eff}$ in this limit. This suppression is due to the fact that the annihilation of two scalars in the $s$-wave corresponds to a $J=0$ state, which implies by angular momentum conservation that the final state must also have $J=0$, which involves both lepton chiralities. Since the Lagrangian from eq.\@ \eqref{eq::lagrangian} involves a chiral interaction,
 the $s$-wave annihilation vanishes in the limit $m_\ell = 0$. The expressions for $a$ and $b$ including the full final-state mass dependence can be found in Appendix \ref{app::partialwave}, and are used in the numerical analysis.
\par 
Note that using equal initial state masses in the equations above is justified in both freeze-out scenarios, as the mass splitting between the flavours $\phi_i$ is negligibly small in the QDF scenario, and the only dark particle assumed to be present during freeze-out in the SFF scenario is the third generation $\phi_3$. Thus, in the SFF scenario $m_\phi$ needs to be replaced by $m_{\phi_3}$ in the expressions above and additionally the sum over initial state flavours and the averaging factor $1/9$ need to be omitted. Setting the final state masses $m_{\ell_i}$ to zero is also justified due to the smallness of the lepton masses and since we generally consider the case $m_\phi \gg m_\tau$.

\subsection{Constraints from the DM Relic Density}
The thermally averaged annihilation cross section that yields the correct DM relic density is found to be independent of the DM mass for $m_\phi > 10\,\mathrm{GeV}$ and reads \cite{Steigman:2012nb, Steigman:2015hda}
\begin{equation}
\langle\sigma v\rangle_\text{eff}^\text{exp} = 2.2 \times 10^{-26}\, \mathrm{cm}^3\,\mathrm{s}^{-1}\,.
\end{equation}
We use the partial wave expansion from eq.\@ \eqref{eq::partialwave} in our numerical analysis and demand that it equals the value from above within a $10\%$ tolerance range. The values for the lepton masses are again adopted from \cite{ParticleDataGroup:2020ssz}.\par
\begin{figure}[t!]
	\centering
		\begin{subfigure}[t]{0.49\textwidth}
		\includegraphics[width=\textwidth]{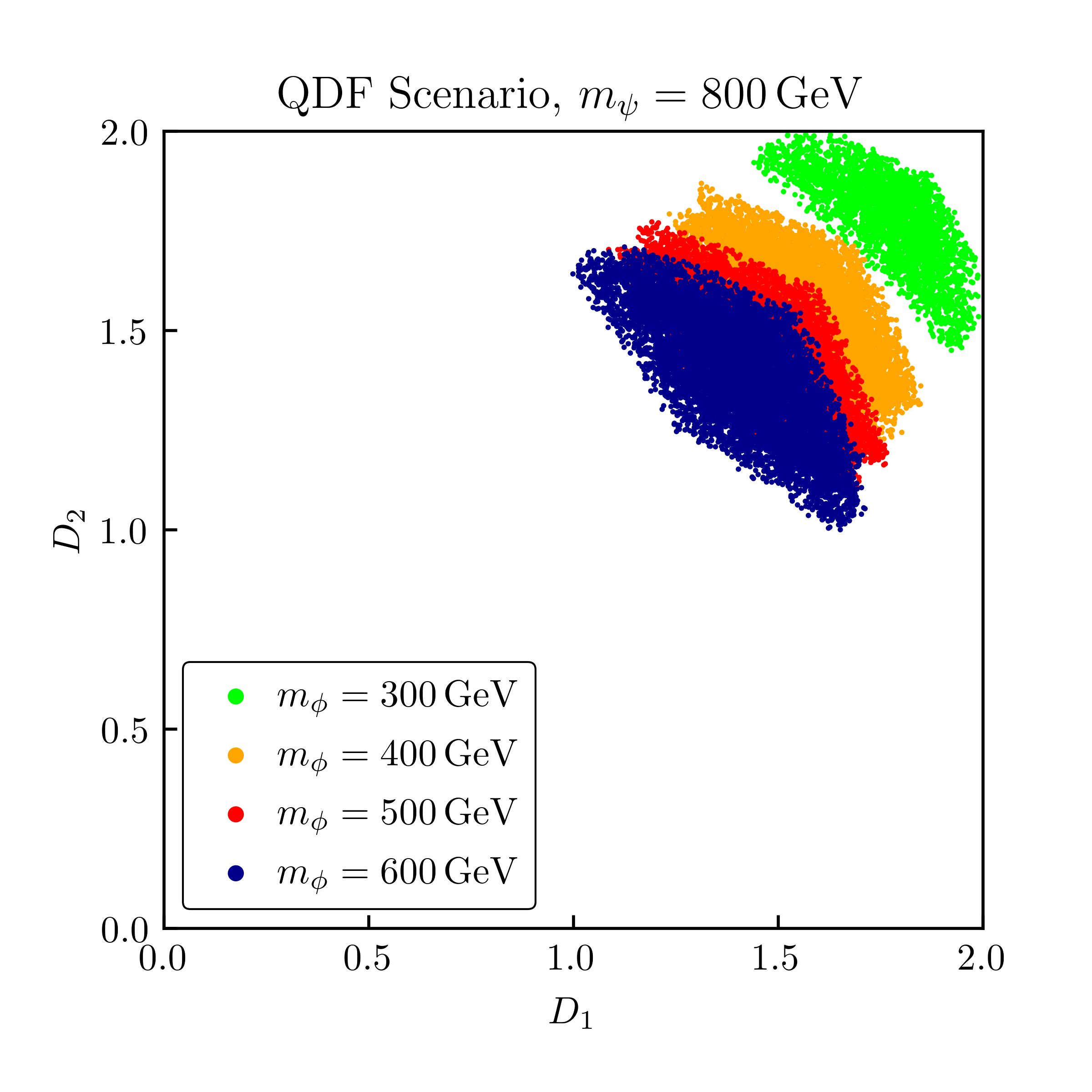}
		\caption{Allowed values of $D_1$ and $D_2$ with a varying $m_\phi$ and $m_\psi = 800\,\mathrm{GeV}$ for $\tau$-flavoured DM in the QDF scenario.}
		\label{fig::relica}
		\end{subfigure}
		\hfill
		\begin{subfigure}[t]{0.49\textwidth}
		\includegraphics[width=\textwidth]{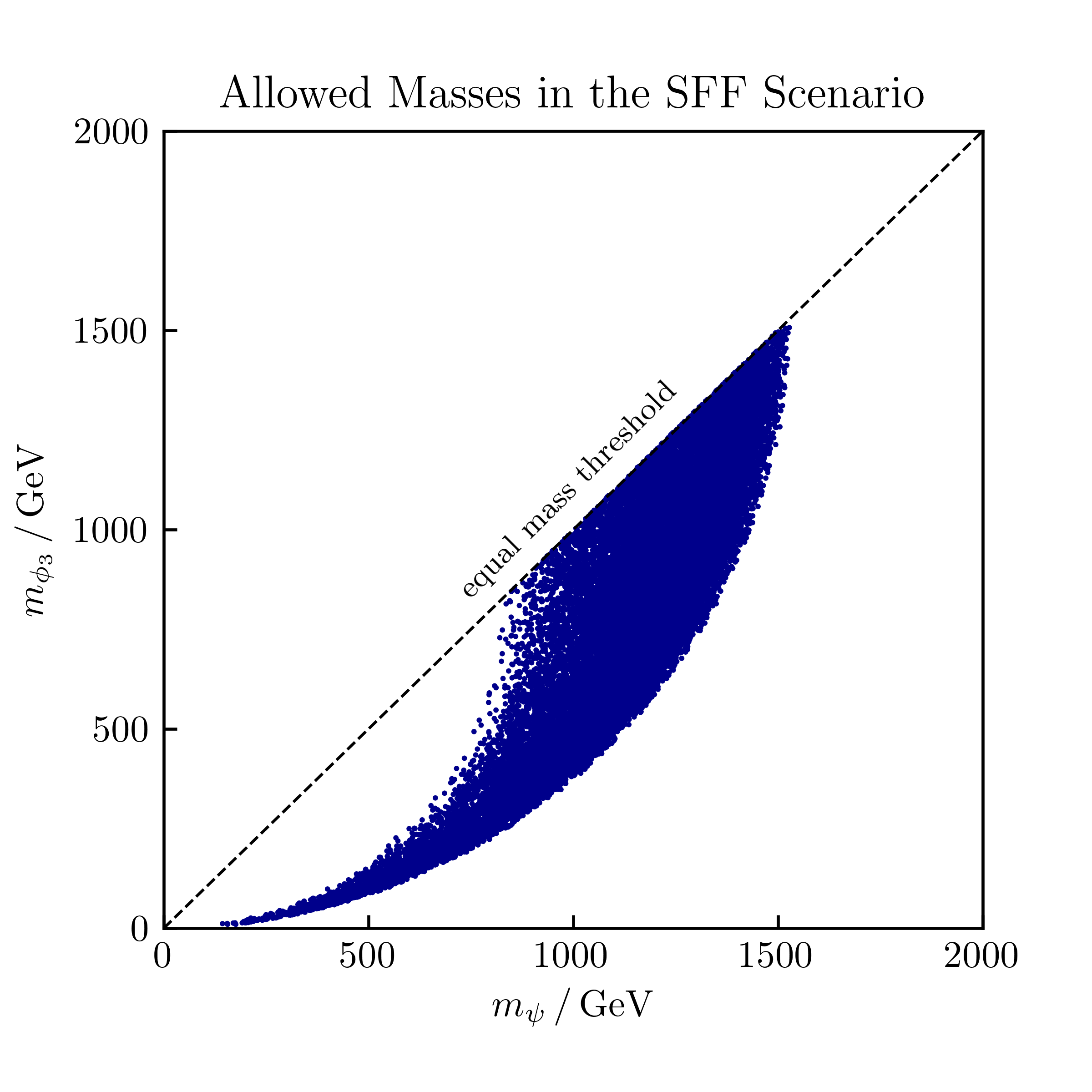}
		\caption{Allowed masses $m_\psi$ and $m_{\phi_3}$ in the SFF scenario.}
		\label{fig::relicb}
		\end{subfigure}
	\caption{Viable parameter space for thermal DM freeze-out.}
	\label{fig::relicconstraints}
	\end{figure}

We show the results of our numerical analysis in Figure \ref{fig::relicconstraints}. 
The contours of Figure \ref{fig::relica} and their mass dependence can both be explained through eq.\@ \eqref{eq::partialb}. Performing the sum over the couplings in the QDF scenario we find
\begin{equation}
b = \frac{1}{432\pi}\frac{m_\phi^2}{(m_\phi^2+m_\psi^2)^2} (D_1^2+D_2^2+D_3^2)^2\,,
\end{equation} 
since we have
\begin{equation}
\sum_{ij}\sum_{kl} |\lambda_{ik}|^2|\lambda_{jl}|^2 = \text{Tr}\left[\lambda^\dagger\lambda\right]^2= \text{Tr}\left[D^2\right]^2\,.
\end{equation}
This means that in the QDF scenario the relic density constraint reduces to a condition on the couplings $D_i$ corresponding to the shell of a sphere. The restrictions that the QDF scenario itself puts on the couplings $D_i$ in order for them to yield the correct mass splittings then further deform this shell and lead to the contours shown in Figure \ref{fig::relica}. As $\langle\sigma v\rangle_\text{eff}$ is proportional to $m_\phi^2$, smaller DM masses require larger couplings in order to yield the correct value for the thermally averaged annihilation cross section. We find that due to the $p$-wave suppression of the latter cross section large couplings $D_i \GtrSim 1.0$ are generally required for viable mediator masses $m_\psi \GtrSim 800\,\mathrm{GeV}$ in order to reproduce the correct relic density.\par
Since the couplings $D_i$ are limited to $D_i \leq 2.0$, the proportionality of the annihilation cross section to $m_\phi^2$ causes a lower limit on the allowed values for $m_\phi$ for a given $m_\psi$ in both scenarios. This can be seen in Figure \ref{fig::relicb} for the SFF scenario. As the latter is defined through a significant mass splitting between the lightest and the heavier states, it further demands that one coupling $D_i$ is significantly larger than the other couplings $D_j$ with $i \neq j$. This is due to the fact that according to eq.\@ \eqref{eq::masssplitting} the mass of a given dark generation $\phi_i$ reads
\begin{equation}
\label{eq::mphii}
m_{\phi_i} = m_\phi \sqrt{1-|\eta|\, D_j^2}\,,
\end{equation}
where the indices $i$ and $j$ are determined by the hierarchy $m_{\phi_1}>m_{\phi_2}>m_{\phi_3}$. Thus the SFF scenario also exhibits an upper limit on $m_{\phi_3}$ for sufficiently small $m_\psi$. This can be seen in Figure \ref{fig::relicb}. Due to one coupling $D_i$ being sizeable $m_{\phi_3}$ can not be chosen arbitrarily large, as this would result in a too large annihilation rate which in turn yields a too small relic density. However, for large $m_\psi \GtrSim 800\,\mathrm{GeV}$ the annihilation rate is sufficiently suppressed by the mediator mass such that values for $m_{\phi_3}$ up to the equal mass threshold $m_{\phi_3} = m_\psi$ become allowed.

\section{DM Detection Experiments}
\label{sec::dmpheno}
One of the key features of lepton-flavoured DM is the absence of tree-level contributions to DM-nucleon scattering which loosens the constraints from direct detection experiments in comparison to quark-flavoured models. However, this feature comes at the cost of rendering indirect detection constraints more relevant, due to the DM particles' direct coupling to electrons and positrons. In this section we discuss both, constraints from direct detection as well as indirect detection experiments.  

\subsection{Direct Detection}
\label{sec::directdetection}
\begin{figure}[t!]
	\centering
		\begin{subfigure}[t]{0.4\textwidth}
		\includegraphics[width=\textwidth]{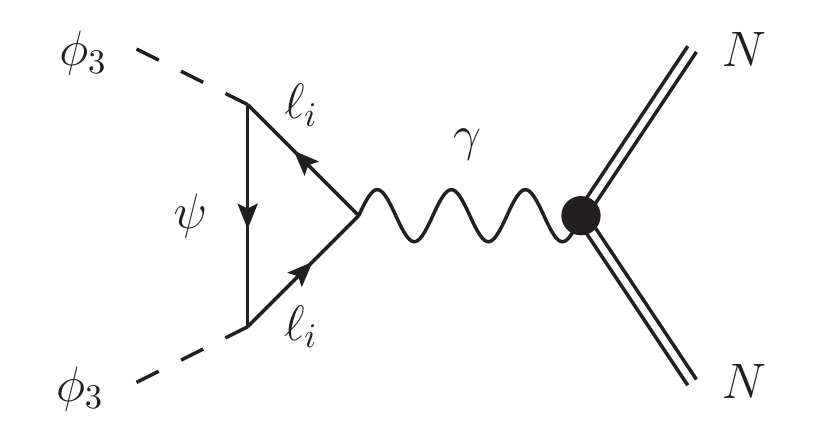}
		\caption{one-loop DM-nucleon scattering}
		\label{fig::photonpenguin}
		\end{subfigure}
		\hspace*{1cm}
		\begin{subfigure}[t]{0.4\textwidth}
		\includegraphics[width=\textwidth]{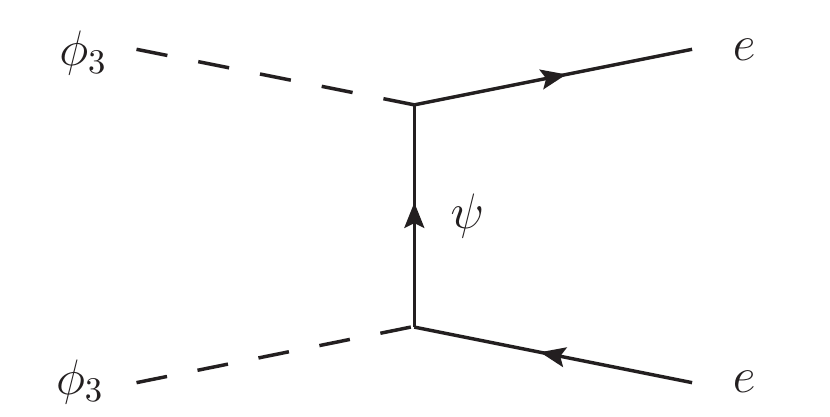}
		\caption{$s$-channel DM-electron scattering}
		\label{fig::electronscattering}
		\end{subfigure}
	\caption{Representative Feynman diagrams of relevant interactions for direct detection signals. Note that for the photon penguin there is also a diagram where the photon is emitted by the mediator $\psi$ and for DM-electron scattering there exists also a $t$-channel diagram.}
	\end{figure}

Lepton-flavoured scalar DM exhibits a very rich direct detection phenomenology as it can give rise to a variety of relevant interactions. Generally, these interactions can be summed up as follows \cite{Kopp:2009et}:
\begin{itemize}
\item[a)] \textit{DM-nucleon scattering}: In this process the DM particle scatters off  nuclei, which results in nuclear recoil signals. This process is generated by  one-loop level diagrams like the one shown in Figure \ref{fig::photonpenguin}. 
\item[b)] \textit{DM-electron scattering}: For DM-electron scattering (see Figure \ref{fig::electronscattering}) we can distinguish between two processes. The first is inelastic scattering, where the DM scatters off  electrons bound in an atom, such that the atom is ionized as the electron absorbs the whole recoil and is kicked out of the atom. The second process is elastic scattering between free electrons and DM in the early universe leading to an inhibition of structure formation and a suppression of cosmic microwave background (CMB) anisotropies. 
\item[c)] \textit{DM-atom scattering}: Here we can again distinguish between inelastic scattering, where DM scatters off  a bound electron which is then excited to an outer shell, and elastic scattering, where the electron wave-function remains the same. In both cases the overall recoil is absorbed by the atom in which the electron is bound.  
\end{itemize}
In spite of this variety of interactions, comparisons between the event rates of DM-nucleon scattering, inelastic DM-electron scattering as well as DM-atom scattering show that scatterings between DM and nuclei strongly dominate over the other two interactions \cite{Kopp:2009et}. In both cases, inelastic DM-electron scattering and DM-atom scattering, it is necessary that DM scatters off  bound electrons with a non-negligible momentum of order $p_e \sim \mathcal{O}(\mathrm{MeV})$ to generate sizeable signals. As the probability for a bound electron to carry such a momentum is small, both interactions suffer from a large wave-function suppression and are thus negligible  compared to DM-nucleon scattering. For elastic DM-electron scattering we on the other hand find  that it only puts constraints on our model for sub-$\mathrm{MeV}$ DM as the limits carry a strong DM mass dependence \cite{Nguyen:2021cnb}.\par  
Following the arguments provided in Section \ref{sec::relicdensity} and in reference \cite{Acaroglu:2021qae}, we  again neglect contributions proportional to the Higgs portal coupling $\lambda_{H\phi}$. Here, the latter gives rise to a tree-level diagram where DM scatters off a nucleon through a $t$-channel Higgs boson exchange. This diagram is proportional to $y_N^2 \lambda_{H\phi}^2$, where $y_N \simeq 0.3$ is the Higgs-nucleon coupling \cite{Cline:2013gha}.\footnote{Note that even without a direct coupling of the DM fields to the Higgs boson an effective coupling is generated at the one-loop level through the same diagram as in Figure \ref{fig::photonpenguin} with a Higgs boson instead of a photon. As these contributions are proportional to $y_{\ell_i} y_N |\lambda_{i3}|^2$, they are negligible due to the smallness of the lepton Yukawa couplings $y_{\ell_i}$.}  Finally, DM also couples to the nucleon's quark vector current. This again happens through the same one-loop diagram as in Figure \ref{fig::photonpenguin} where in this case instead of a photon a $Z$-boson mediates the scattering process. The amplitude of such a process is proportional to $m_{\ell_i}^2/m_\psi^2$ \cite{Kawamura:2020qxo}, where $m_{\ell_i}$ is the mass of the lepton in the loop. We hence also neglect contributions from $Z$-boson penguin diagrams.\par  
In summary, for the analysis of constraints that direct detection experiments put on our model we focus on the one-loop scattering between DM and nucleons shown in Figure \ref{fig::photonpenguin}. This contribution is induced by the charge-radius operator
\begin{equation}
\mathcal{O}_\gamma = \partial^\mu \phi \,\partial^\nu \phi^\dagger F_{\mu\nu}\,.
\end{equation}
In the limit of negligible lepton masses, i.e.\@ $m_{\ell_i} \ll m_\psi$, its matched Wilson coefficient $f_\gamma$ reads \cite{Bai:2014osa}
\begin{equation}
\label{eq::fgamma}
f_\gamma = -\sum_i \frac{e\,|\lambda_{i3}|^2}{16\pi^2 \,m_\psi^2}  \left[1+ \frac{2}{3} \ln\left(\frac{m_{\ell_i}^2}{m_\psi^2}\right) \right]\,.
\end{equation}
Note that for $i=1$, i.e.\@ for an electron-positron pair in the loop, the mass $m_e$ needs to be replaced by the momentum transfer $|\vec{q}|=\mathcal{O}(3-10)\,\mathrm{MeV}$ as the electron mass is smaller than $|\vec{q}|$ \cite{Bai:2014osa}. 

For the spin-independent averaged DM-nucleon scattering cross section we find 
\begin{equation}
\label{eq::SIcx}
\sigma^N_\text{SI} = \frac{Z^2\, e^2\, \mu^2}{8\pi\,A^2}\,f_\gamma^2\,,
\end{equation}
where $Z$ and $A$ are the atomic and mass number of the element that the nucleons constitute. Here, $\mu$ is the reduced mass of the DM-nucleon system defined as $\mu = m_N m_{\phi_3} /(m_N + m_{\phi_3})$. Using this cross section together with limits obtained from the XENON1T experiment \cite{XENON:2018voc} we  constrain our model numerically. We again use the lepton masses obtained from \cite{ParticleDataGroup:2020ssz} and set $|\vec{q}| = 10\, \mathrm{MeV}$ for the momentum transfer mentioned above. The atomic and mass numbers of Xenon read $Z = 54$ and $A = 131$.\par
\begin{figure}[b!]
	\centering
		\begin{subfigure}[t]{0.49\textwidth}
		\includegraphics[width=\textwidth]{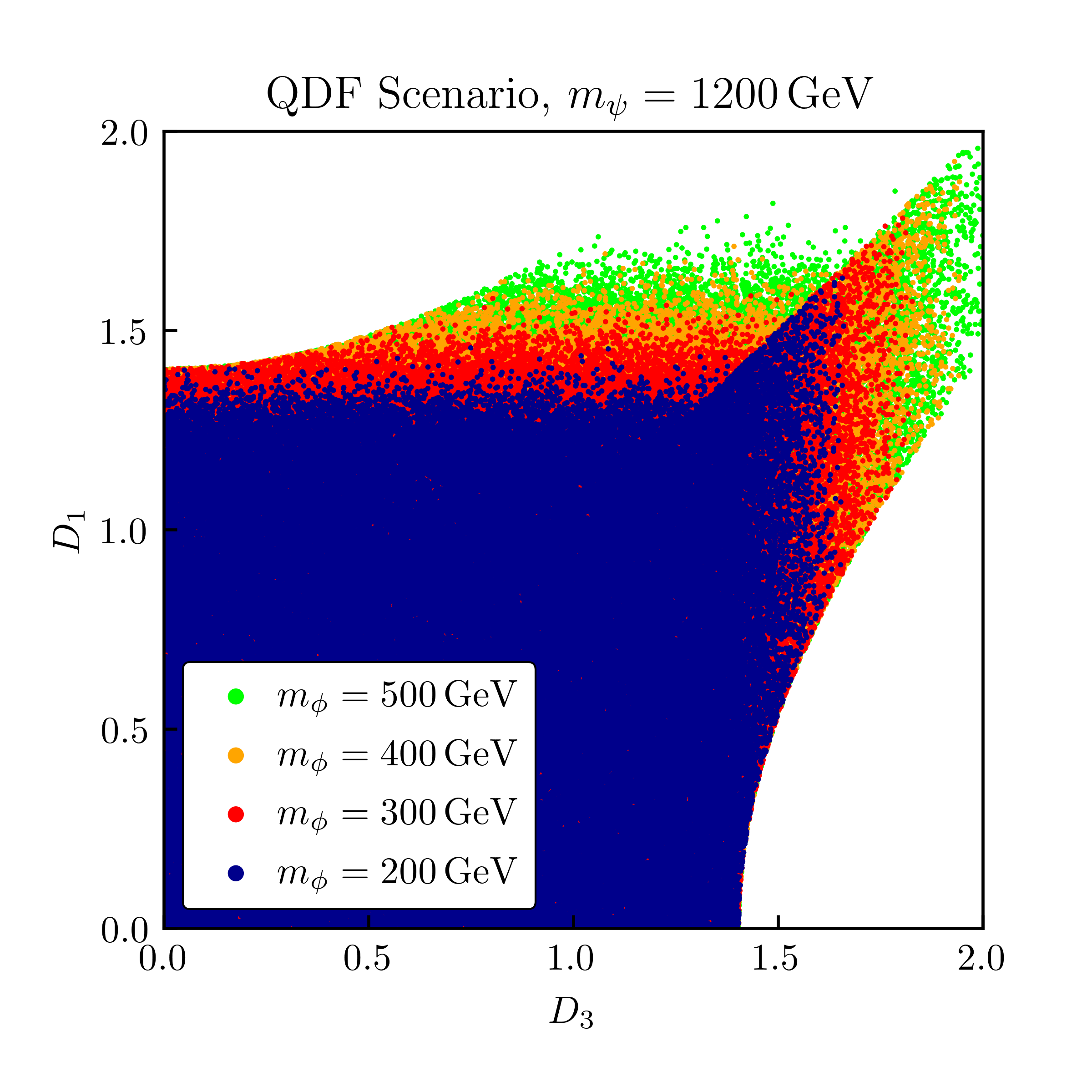}
		\caption{QDF scenario}
		\label{fig::directa}
		\end{subfigure}
		\hfill
		\begin{subfigure}[t]{0.49\textwidth}
		\includegraphics[width=\textwidth]{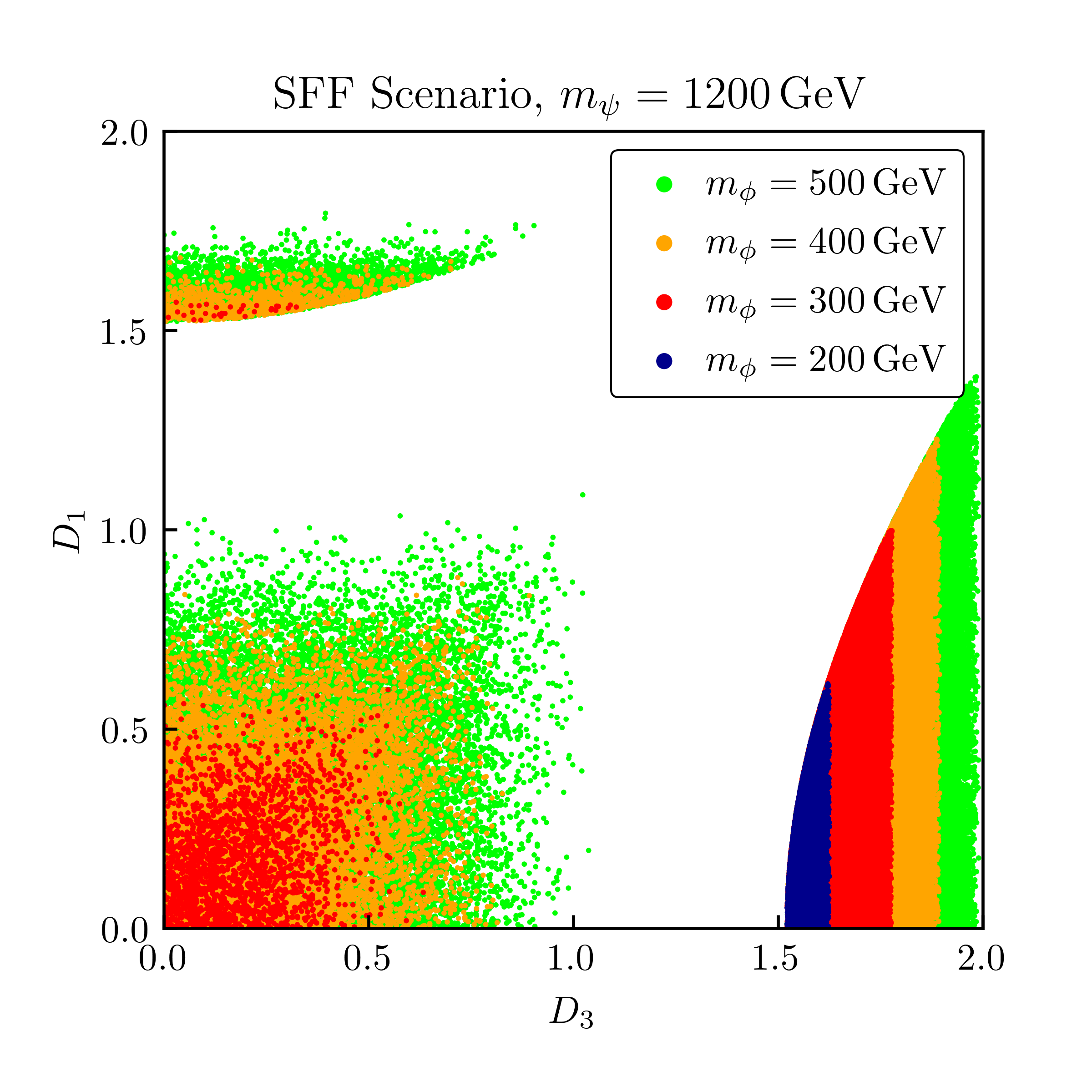}
		\caption{SFF scenario}
		\label{fig::directb}
		\end{subfigure}
	\caption{Allowed couplings $D_3$ and $D_1$ in both scenarios. The mediator mass is fixed to $m_\psi = 1200\,\mathrm{GeV}$ and the tree-level DM mass parameter $m_\phi$ varies.}
	\label{fig::directconstraints}
	\end{figure}

  
The results are shown in Figure \ref{fig::directconstraints}. 
In both scenarios $\tau$-flavoured DM, i.e.\@ points in the parameter space with $D_3 > D_{1,2}$ generally allow for larger couplings $D_i$ than $e$- or $\mu$-flavoured DM. This is due to the $m_\ell$ dependence of the photon penguin diagram from Figure \ref{fig::photonpenguin}. Its amplitude from eq.\@ \eqref{eq::fgamma} is proportional to the logarithm of the mass of the lepton in the loop and thus it grows with a decreasing $m_\ell$. Hence, we find the largest restrictions on $D_1$ and $D_2$ as the tau mass $m_\tau$ is much larger than the muon mass $m_\mu$ or the momentum transfer $|\vec{q}|$. While the amplitude $f_\gamma$ exhibits a $1/m_\psi^2$ suppresion, it does not depend on the DM mass $m_{\phi_3}$. This parameter only enters the calculation through the reduced mass $\mu$, for which we find $\mu \approx m_N$ since $m_{\phi_3} \gg m_N$. Thus, the only $m_{\phi_3}$ dependence comes from the XENON1T upper limit itself, which reaches its minimum value for $m_{\phi_3} \simeq 30\,\mathrm{GeV}$ and grows with increasing DM mass values. This explains why in both scenarios larger values $D_i$ are allowed for increasing values of $m_{\phi_3}$.\par
The points above the diagonal in Figure \ref{fig::directa} correspond to either $\mu$- or $e$-flavoured DM, while the points below the diagonal represent either $\mu$- or $\tau$-flavoured DM. We see that for the reasons explained above the latter allows for larger couplings $D_i$. In the SFF scenario shown in Figure \ref{fig::directb} the points with a large $D_3$ correspond to $\tau$-flavour, the points with a large $D_1$ to $e$-flavour and the points close to the origin to $\mu$-flavour. As the SFF scenario forces one coupling $D_i$ to be significantly larger than the others, we find that for $e$- and $\mu$-flavoured DM the direct detection constraints cannot be fulfilled for small DM masses, $m_{\phi_3} = 200\,\mathrm{GeV}$. This is due to the enhancement of $f_\gamma$ for $e$- and $\mu$-flavoured DM and the more stringent experimental upper limit for smaller mass $m_{\phi_3}$.

\subsection{Indirect Detection}
\label{sec::indirectdetection}
As already mentioned in the introduction of this section, lepton-flavoured DM can generally lead to large signals for indirect detection experiments. This is due to the fact that the direct coupling of $\phi_3$ to electrons can lead to sizeable electron-positron fluxes. However, due to the $p$-wave suppression of the thermally averaged annihilation cross section discussed in Section \ref{sec::relicdensity}, this is not the case for lepton-flavoured scalar DM. The DM halo velocity in the Milky Way today is much smaller than the relative velocity during the freeze-out and reads $\langle v^2 \rangle \simeq 10^{-6}$. Hence, $\langle\sigma v\rangle_\text{eff}$ suffers from a severe velocity suppression.\par
\begin{figure}[t!]
	\centering
		\begin{subfigure}[t]{0.4\textwidth}
		\includegraphics[width=\textwidth]{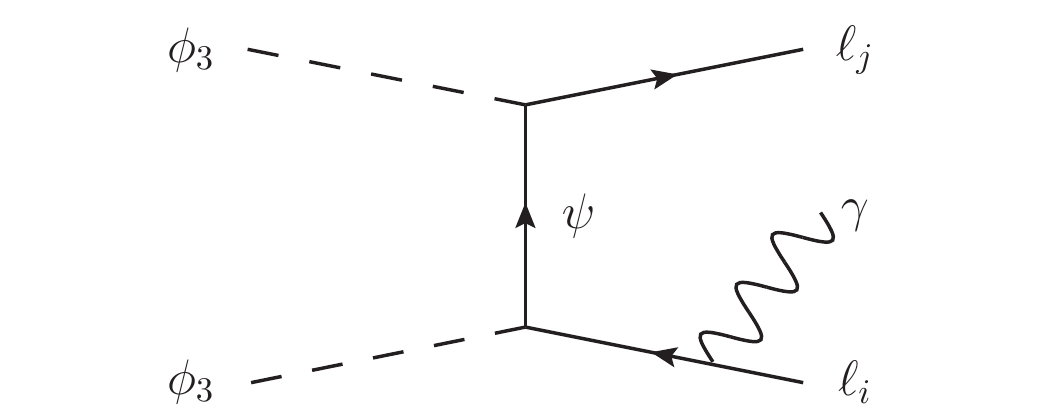}
		\caption{$t$-channel annihilation to $\ell\bar{\ell}\gamma$}
		\label{fig::iddiag1}
		\end{subfigure}
		\hspace*{1cm}
		\begin{subfigure}[t]{0.4\textwidth}
		\includegraphics[width=\textwidth]{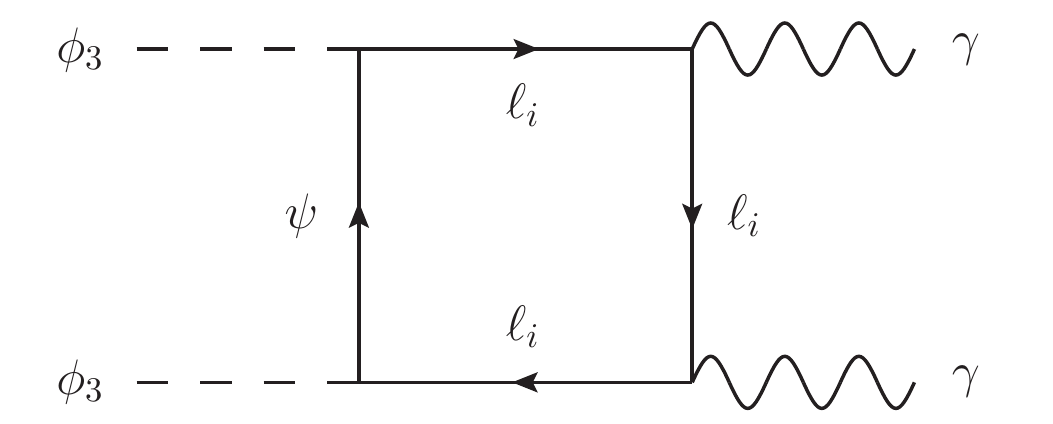}
		\caption{one-loop annihilation to $\gamma\gamma$}
		\label{fig::iddiag2}
		\end{subfigure}
	\caption{Representative Feynman diagrams for relevant higher-order annihilation processes.}
	\label{fig::iddiag}
	\end{figure}

 
It is thus necessary to include the additional diagrams shown in Figure \ref{fig::iddiag} to the calculation of $\langle\sigma v\rangle_\text{eff}$ to provide a proper analysis of the constraints that indirect detection places on our model. 
The process of two dark particles $\phi_3$ annihilating into the three-body final state $\ell_i \bar{\ell}_j \gamma$ shown in Figure \ref{fig::iddiag1} lifts the $p$-wave suppression of the annihilation rate. The one-loop annihilation into two photons shown in Figure \ref{fig::iddiag2} on the other hand gives comparably sizeable contributions in spite of its loop suppression. Note that both diagrams are not relevant for the thermal freeze-out as the one-loop diagram is parametrically suppressed by $\alpha_\text{em}^2/(4\pi)^2 \sim 10^{-7}$ while the diagram with a single photon in the final state is suppressed by $\alpha_\text{em}/\pi \sim 10^{-3}$. The annihilation rate into $\ell_i \bar \ell_j$ in contrast is only suppressed by $\langle v^2\rangle \simeq 0.3$ in the early universe. \par 
The annihilation rates $\langle\sigma v\rangle_{\ell\bar{\ell}\gamma}$ and $\langle\sigma v\rangle_{\gamma\gamma}$  for the processes shown in Figure \ref{fig::iddiag} in the limit of vanishing lepton masses $m_\ell \rightarrow 0$ read \cite{Toma:2013bka, Tulin:2012uq}
\begin{align}
\label{eq::idllgamma}
\langle\sigma v\rangle_{\ell\bar{\ell}\gamma} &= \frac{\alpha_\text{em}}{32 \pi^2 m_{\phi_3}^2}\sum_{ij} |\lambda_{i3}|^2 |\lambda_{j3}|^2 \mathcal{A(\mu)}\,,\\
\label{eq::idgammagamma}
\langle\sigma v\rangle_{\gamma\gamma} &= \frac{\alpha_\text{em}^2}{64 \pi^3 m_{\phi_3}^2}\left(\sum_i |\lambda_{i3}|^2\right)^2 |\mathcal{B(\mu)}|^2\,.
\end{align}
The functions $\mathcal{A}$ and $\mathcal{B}$ are defined as
\begin{align}
\nonumber
\mathcal{A}(\mu) &= (\mu+1) \left(\frac{\pi^2}{6}-\log^2\left[\frac{\mu+1}{2\mu}\right]-2\text{Li}_2\left[\frac{\mu+1}{2\mu}\right]\right)\\
&\phantom{=}+\frac{4\mu +3}{\mu +1}+\frac{4\mu^2-3\mu-1}{2\mu}\log\left[\frac{\mu-1}{\mu+1}\right]\,,\\
\mathcal{B}(\mu) &= 2-2 \log\left[1-\frac{1}{\mu}\right]-2 \mu \arcsin\left[\frac{1}{\sqrt{\mu}}\right]^2 \,,
\end{align}
where $\mu = m_\psi^2/m_{\phi_3}^2$ and $\text{Li}_2(z)$ is the dilogarithm. The rate for the tree-level annihilation into a pair of leptons is the same as the thermal annihilation cross section for the SFF scenario discussed in Section \ref{sec::relicdensity}.\par

For our numerical analysis we use limits based on measurements of the AMS \cite{AMS:2013fma}, Fermi-LAT \cite{Bringmann:2012vr} and H.E.S.S. \cite{HESS:2013rld} experiments to constrain the parameter space of our model. Using AMS-02 measurements of the positron flux, reference \cite{Ibarra:2013zia} provides an upper limit $\langle\sigma v\rangle_{\bar{e}}^\text{max}$ on the annihilation rate of DM to an electron-positron pair with a branching ratio of 100$\%$. The positron flux signal generally includes both, prompt as well as secondary positrons from decays of muons and taus. However, since the energy spectrum is shifted towards lower energies for the latter and since they additionally suffer from a smeared momentum distribution, the signal is mainly dominated by prompt positrons. Thus, for the numerical analysis we sum over all annihilation channels with a positron in the final state  and compare with the experimental upper limit. Since we expect the shift in the DM mass dependence for the three-body final state compared to the limit calculated for a two-body final state in \cite{Ibarra:2013zia} to be negligible, we here include the radiative corrections of Figure \ref{fig::iddiag1}, i.e.\@ we demand that the sum
\begin{equation}
\label{eq::idsigmae}
\langle\sigma v\rangle_{\bar{e}} = \sum_{\ell}\langle\sigma v\rangle_{\ell\bar{e}}+\langle\sigma v\rangle_{\ell\bar{e}\gamma}\,,
\end{equation} 
is smaller than the experimental upper limit $\langle\sigma v\rangle_{\bar{e}}^\text{max}$.\par
In \cite{Tavakoli:2013zva} an upper limit $\langle\sigma v\rangle_{\tau}^\text{max}$ is provided for annihilations into a tau-antitau pair in a similar fashion based on measurements of the $\gamma$-ray continuum spectrum by Fermi-LAT. Similar to the case of the positron flux being most sensitive to prompt positrons, this signal is mainly dominated by taus or antitaus in the final state as they produce more photons through subsequent decays than electrons and muons. We thus calculate the total annihilation rate into final states with at least one tau or antitau, i.e.\@ we calculate the sum
\begin{align}
    \langle\sigma v\rangle_{\tau} &= \langle\sigma v\rangle_{\tau\bar{\tau}}+\langle\sigma v\rangle_{\tau\bar{\tau}\gamma}+\frac{1}{2}\sum_{\ell=e,\mu}\left(\langle\sigma v\rangle_{\ell\bar{\tau}}+\langle\sigma v\rangle_{\bar{\ell}\tau}+\langle\sigma v\rangle_{\ell\bar{\tau}\gamma}+\langle\sigma v\rangle_{\bar{\ell}\tau\gamma}\right)\,,
    \label{eq::idsigmatau}
\end{align}
and compare it with the upper limit $\langle\sigma v\rangle_{\tau}^\text{max}$. Here we have again additionally included annihilations into the three-body final state and a factor of $1/2$ for final states with a single tau or antitau, since the respective upper limit was derived for a tau-antitau pair in the final state.\par 
Finally, reference \cite{Garny:2013ama} uses H.E.S.S. and Fermi-LAT data of the $\gamma$-ray line spectrum to provide an upper limit $\langle\sigma v\rangle_\gamma^\text{max}$ on the sum
\begin{equation}
\label{eq::idsigmagamma}
\langle\sigma v\rangle_\gamma=\langle\sigma v\rangle_{\ell\bar{\ell}\gamma} + 2\langle\sigma v\rangle_{\gamma\gamma}\,,
\end{equation}
since both processes of Figure \ref{fig::iddiag} exhibit a line-like photon energy spectrum. 
For the $2 \to 2$ process induced by the box diagram of Figure \ref{fig::iddiag2} the energy distribution peaks at $E_\gamma = m_{\phi_3}$. The process depicted in Figure \ref{fig::iddiag1} on the other hand is a three-body process, but resembles a line signal because the internal bremsstrahlung photons emitted from the virtual $\psi$ exhibit a sharply peaked spectrum just below the DM mass \cite{Garny:2013ama}.\par 
In order to estimate the impact of the constraints on our model, we determine the coupling strength for which each of the annihilation rates $\langle\sigma v\rangle_{\bar{e}}$, $\langle\sigma v\rangle_\tau$ and $\langle\sigma v\rangle_\gamma$ saturates its respective experimental upper limit. To this end, we scan over the mass parameters $m_\psi$ and $m_{\phi_3}$ and take the limit of degenerate couplings $|\lambda_{i3}| = |\lambda_{\ell3}|$ when calculating the rates of eqs.\@ \eqref{eq::idsigmae} -- \eqref{eq::idsigmagamma}.\par
\begin{figure}[b!]
	\centering
		\begin{subfigure}[t]{0.49\textwidth}
		\includegraphics[width=\textwidth]{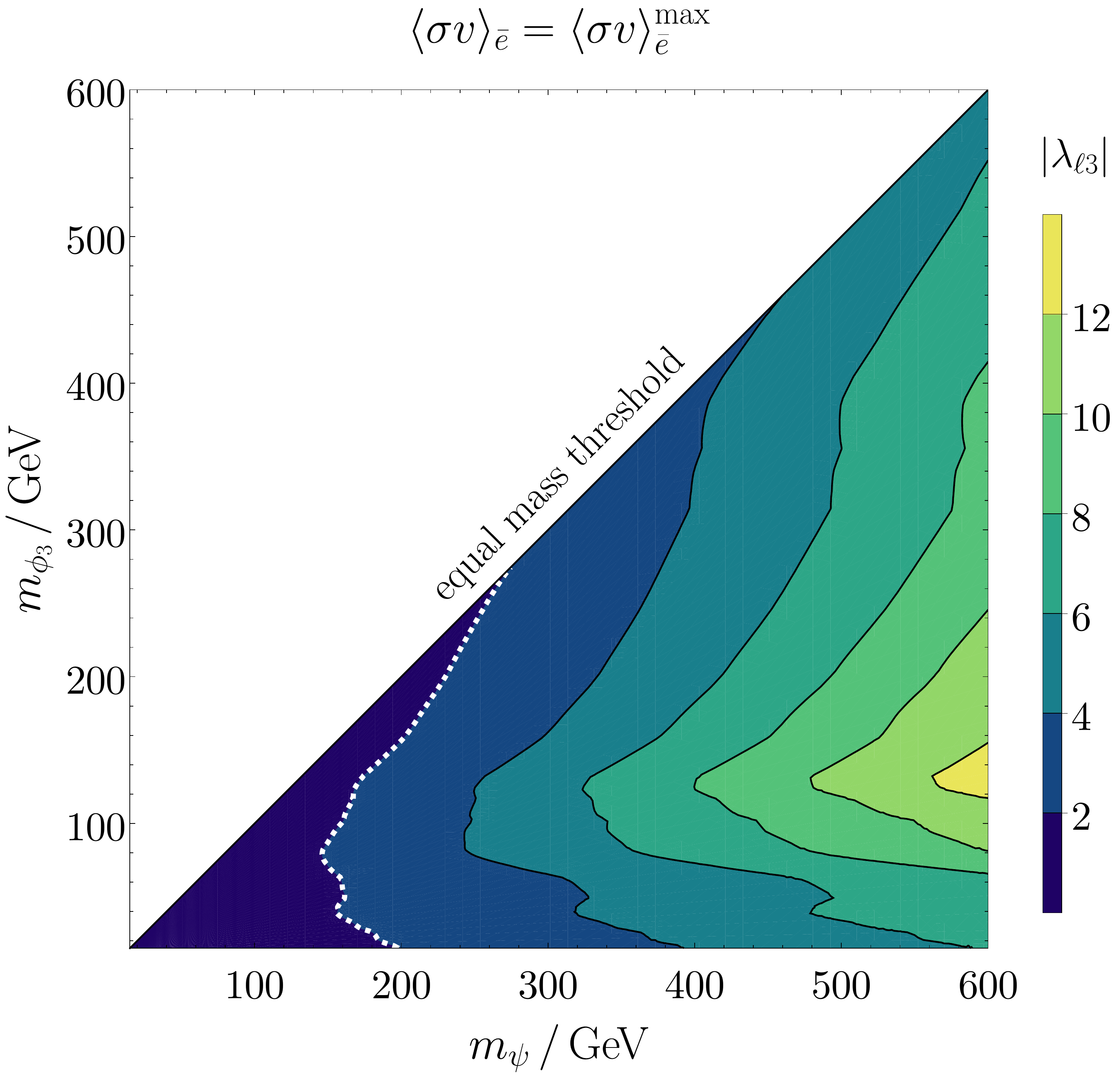}
		\caption{constraints from the positron flux}
		\label{fig::idelec}
		\end{subfigure}
		\hfill
		\begin{subfigure}[t]{0.49\textwidth}
		\includegraphics[width=\textwidth]{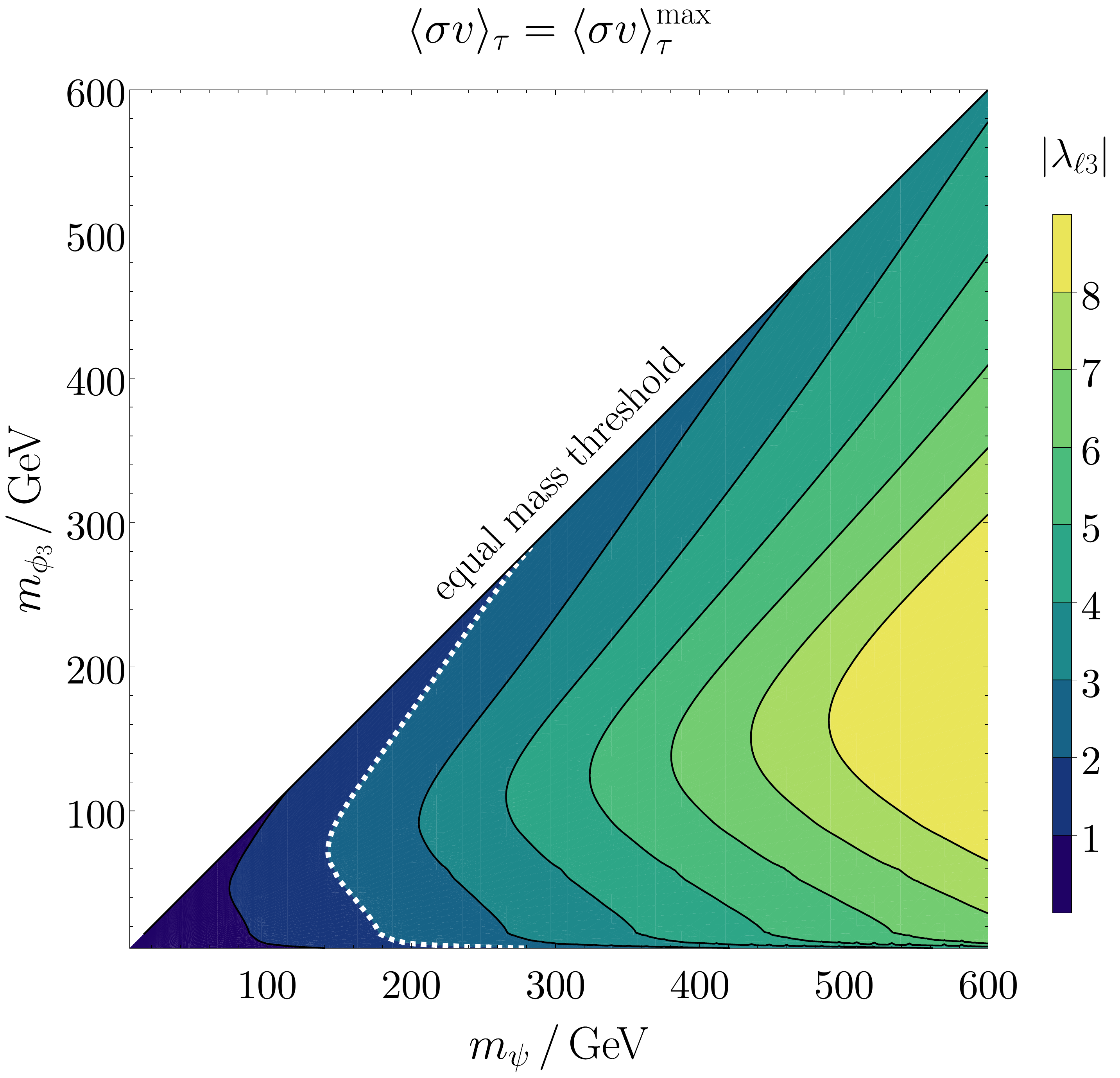}
		\caption{constraints from the $\gamma$ continuum spectrum}
		\label{fig::idtau}
		\end{subfigure}
		\hfill
		\begin{subfigure}[t]{0.49\textwidth}
		\vspace{0.2cm}
		\includegraphics[width=\textwidth]{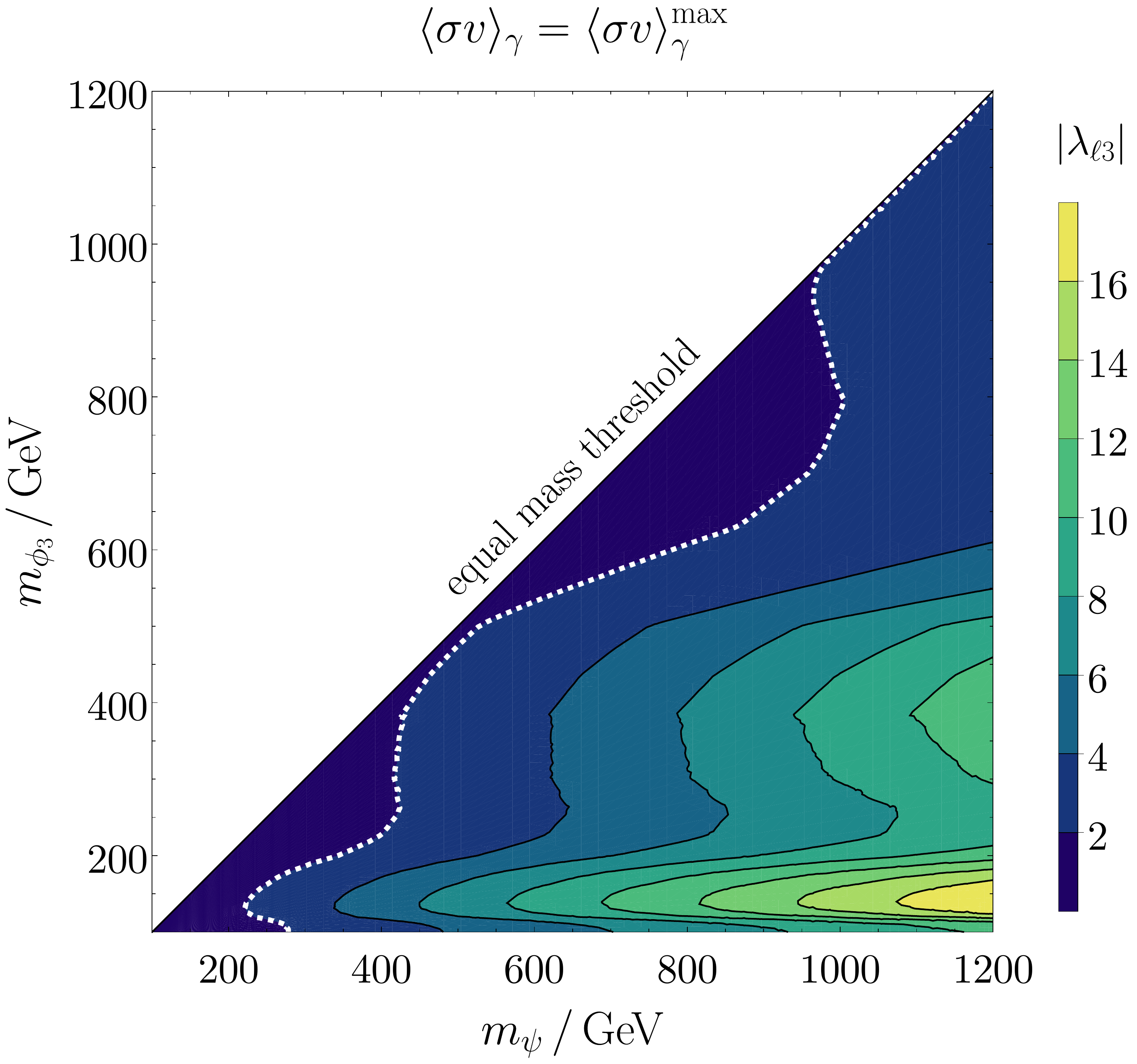}
		\caption{constraints from the $\gamma$ line spectrum}
		\label{fig::idgamma}
		\end{subfigure}
	\caption{Restrictions on the model parameters from indirect detection experiments. The area included by the white dashed line and the equal mass diagonal indicates in which mass regime the constraints are relevant.}
	\label{fig::idconst}
	\end{figure}

  
The results are shown in Figure \ref{fig::idconst}. In all three panels the white dashed line indicates the contour with $|\lambda_{i3}| = 2.0$.
As the couplings are limited to $|\lambda_{i3}| \in [0, 2]$, we only expect constraints on our model in the area that is enclosed by the equal mass diagonal and this contour. In Figure \ref{fig::idelec} we see that limits obtained from measurements of the positron flux or the $\gamma$-ray continuum spectrum only lead to exclusions in our parameter space for masses $m_\psi$ which are already excluded by the LHC searches discussed in Section \ref{sec::collider}. The limit for annihilations into a single positron is relevant for $m_\psi \LessSim 250\,\mathrm{GeV}$, where between $200\,\mathrm{GeV} \LessSim m_\psi \LessSim 250\,\mathrm{GeV}$ it only constrains our model in the region close to the degeneracy limit $m_\psi = m_{\phi_3}$. For annihilations into final states with a tau or antitau we find a similar behaviour, with the sole difference that in this case limits can be relevant up to mediator masses $m_\psi \approx 300\,\mathrm{GeV}$. In Figure \ref{fig::idgamma} we see that measurements of the $\gamma$-ray line spectrum place more stringent constraints on the coupling matrix $\lambda$. Here the limits are mainly relevant in the near-degeneracy regime $m_\psi \approx m_{\phi_3}$, and we find that they put constraints on $\lambda$ for masses up to $m_\psi \simeq 1000\,\mathrm{GeV}$. In total, we conclude that the indirect detection constraints remain rather weak when compared to limits from direct detection experiments or the observed DM relic density.
\section{Combined Analysis}
\label{sec::combined}
\begin{figure}[b!]
	\centering
		\includegraphics[width=0.49\textwidth]{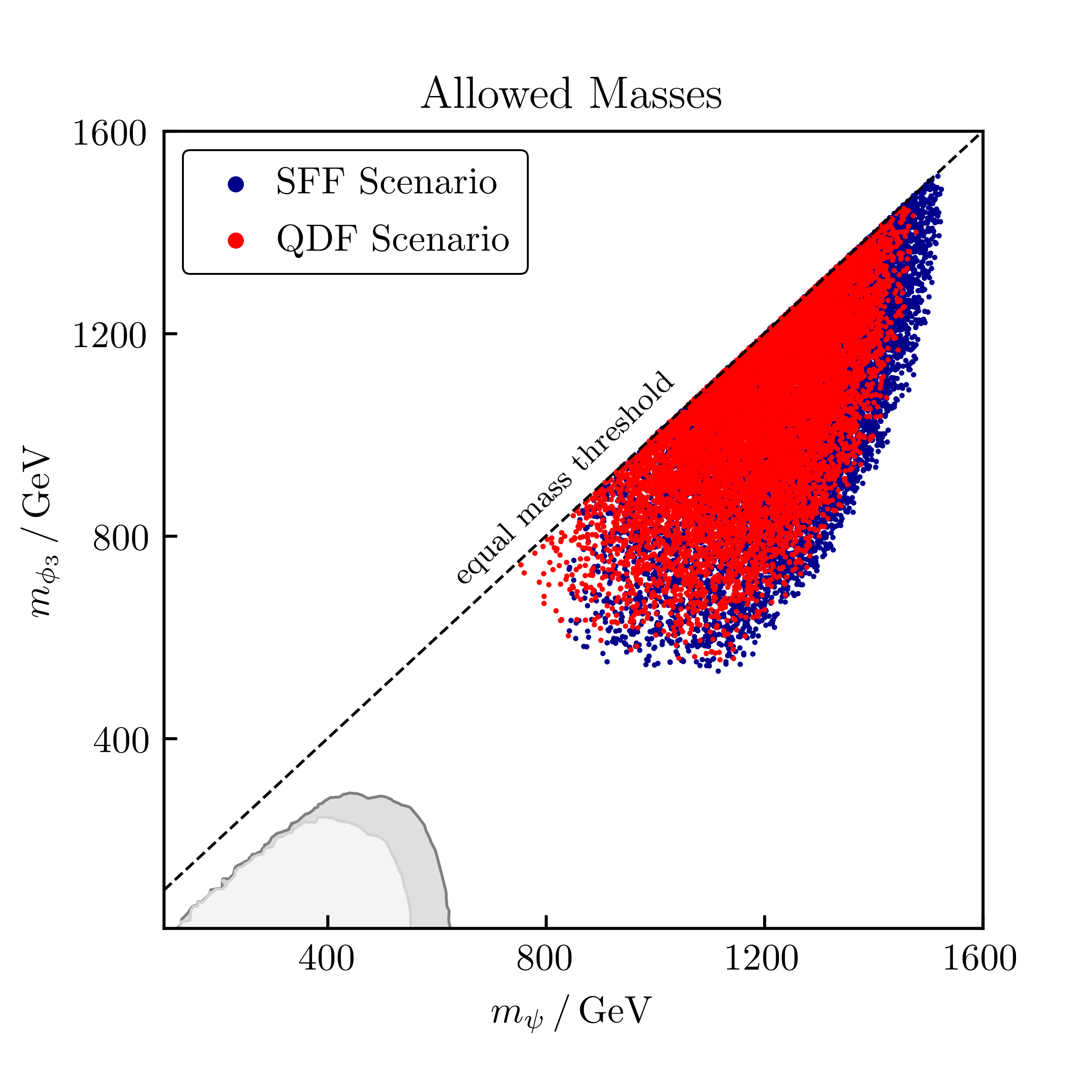}
	\caption{Allowed masses in both scenarios when fulfilling all constraints at the $2\sigma$ level. We further show the largest possible exclusion contours coming from the LHC searches for same-flavour final states $\ell \bar{\ell} + \slashed{E}_T$ with $\ell = e, \mu$ in both scenarios. The grey contour corresponds to the QDF scenario with $D_1 = D_2 = D_3 = 2.0$. The light-grey contour represents the largest possible exclusion in the SFF scenario with $D_1 = D_2 = 1.5$ and $D_3=2.0$.}
	\label{fig::cbdmasses}
	\end{figure}

\begin{figure}[b!]
	\centering
		\begin{subfigure}[t]{0.49\textwidth}
		\includegraphics[width=\textwidth]{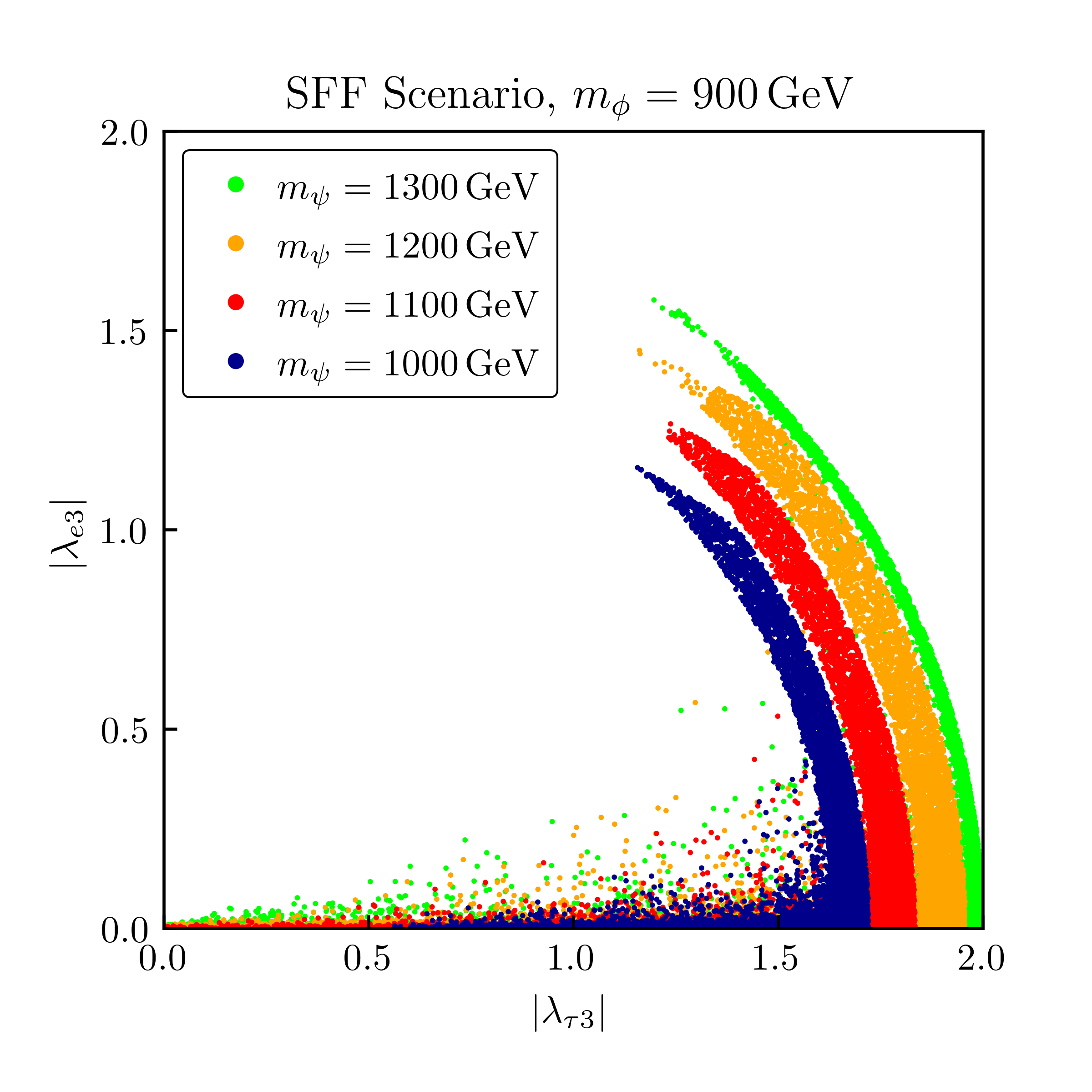}
		\caption{$|\lambda_{\tau 3}| - |\lambda_{e 3}|$ plane}
		\label{fig::cbdsffa}
		\end{subfigure}
		\hfill
		\begin{subfigure}[t]{0.49\textwidth}
		\includegraphics[width=\textwidth]{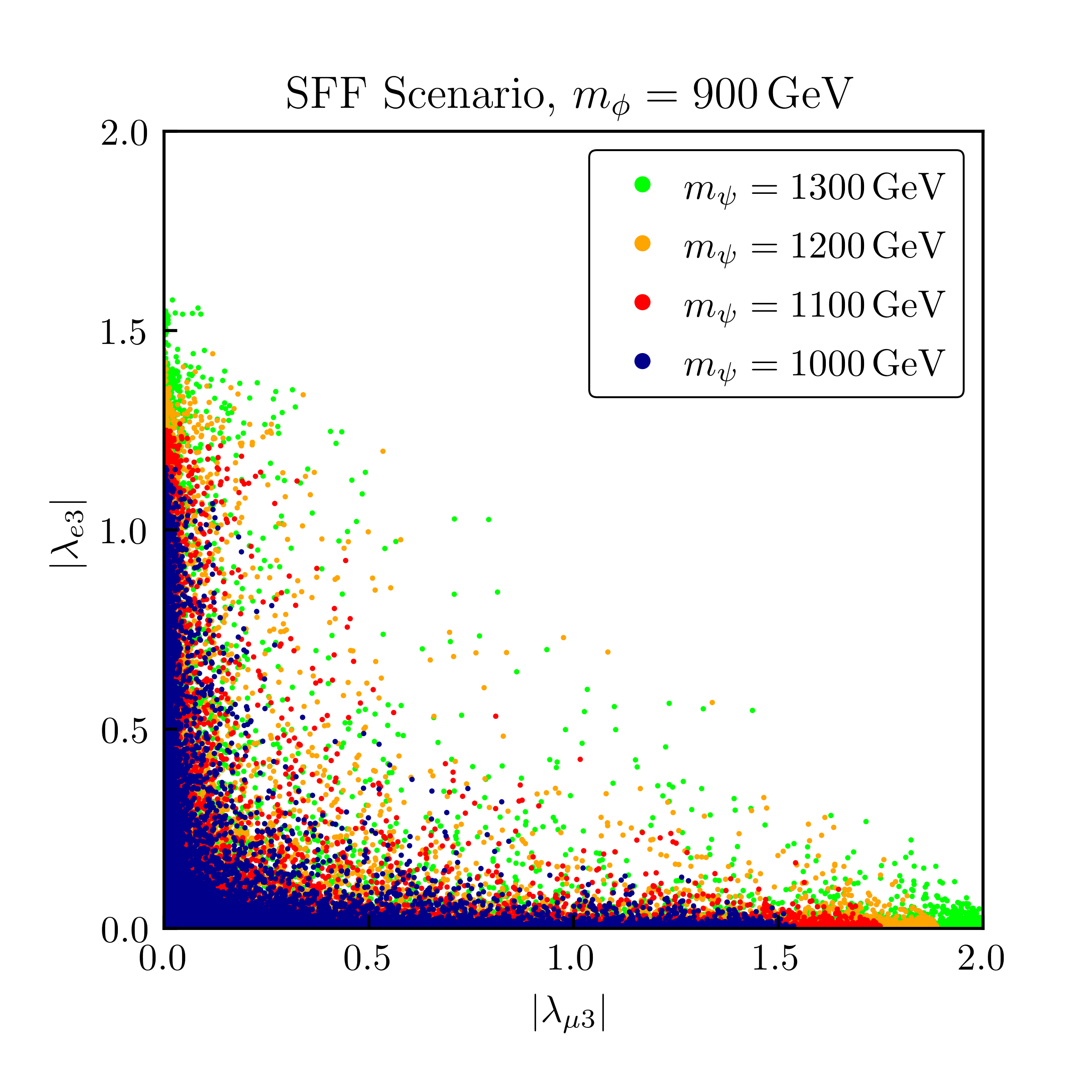}
		\caption{$|\lambda_{\mu 3}| - |\lambda_{e 3}|$ plane}
		\label{fig::cbdsffb}
		\end{subfigure}
	\caption{Allowed couplings $|\lambda_{i3}|$ for $m_\phi = 900\,\mathrm{GeV}$ and varying $m_\psi$ based on all constraints in the SFF scenario.}
	\label{fig::cbdsff}
	\end{figure}

\begin{figure}[b!]
	\centering
		\begin{subfigure}[t]{0.49\textwidth}
		\includegraphics[width=\textwidth]{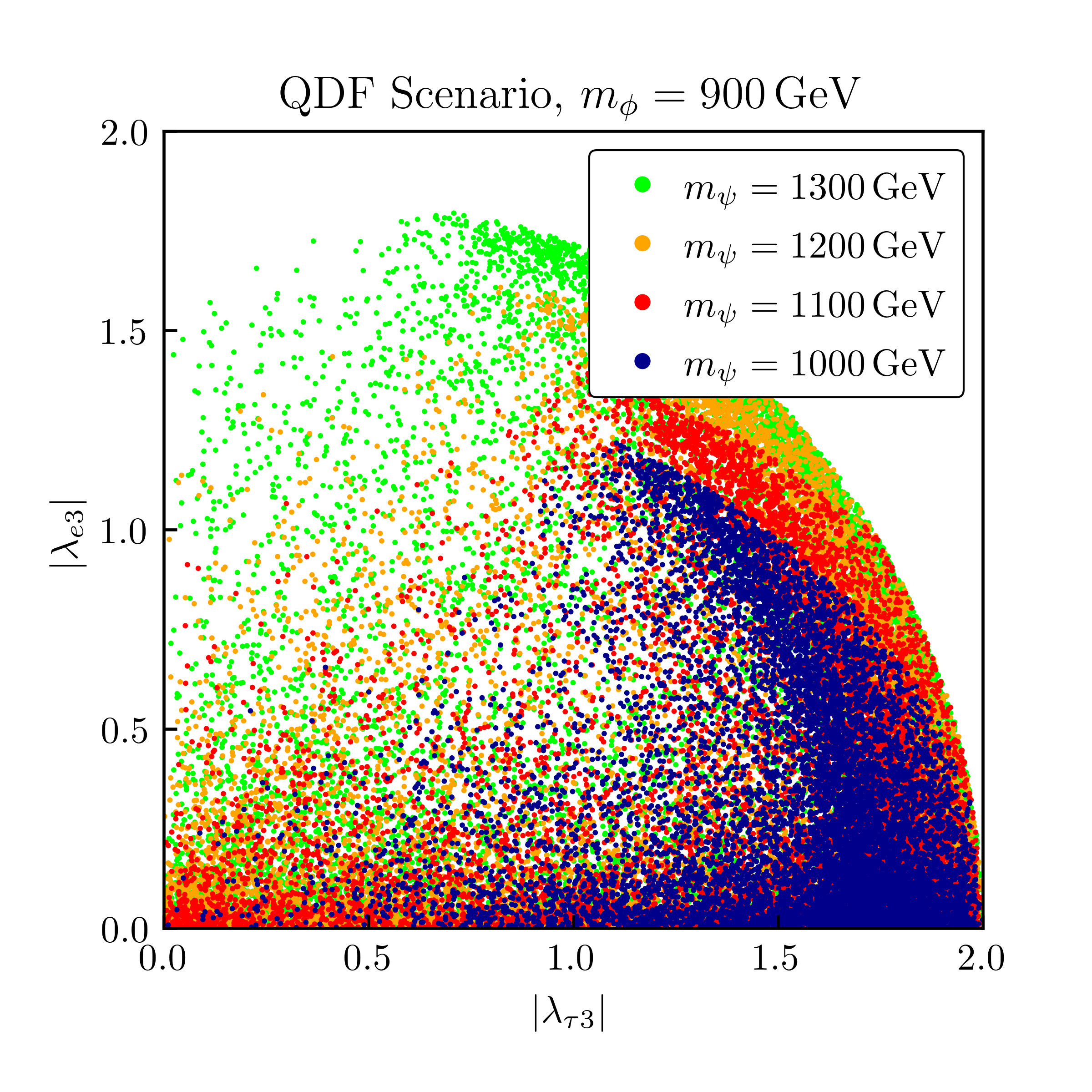}
		\caption{$|\lambda_{\tau 3}| - |\lambda_{e 3}|$ plane}
		\label{fig::cbdqdfa}
		\end{subfigure}
		\hfill
		\begin{subfigure}[t]{0.49\textwidth}
		\includegraphics[width=\textwidth]{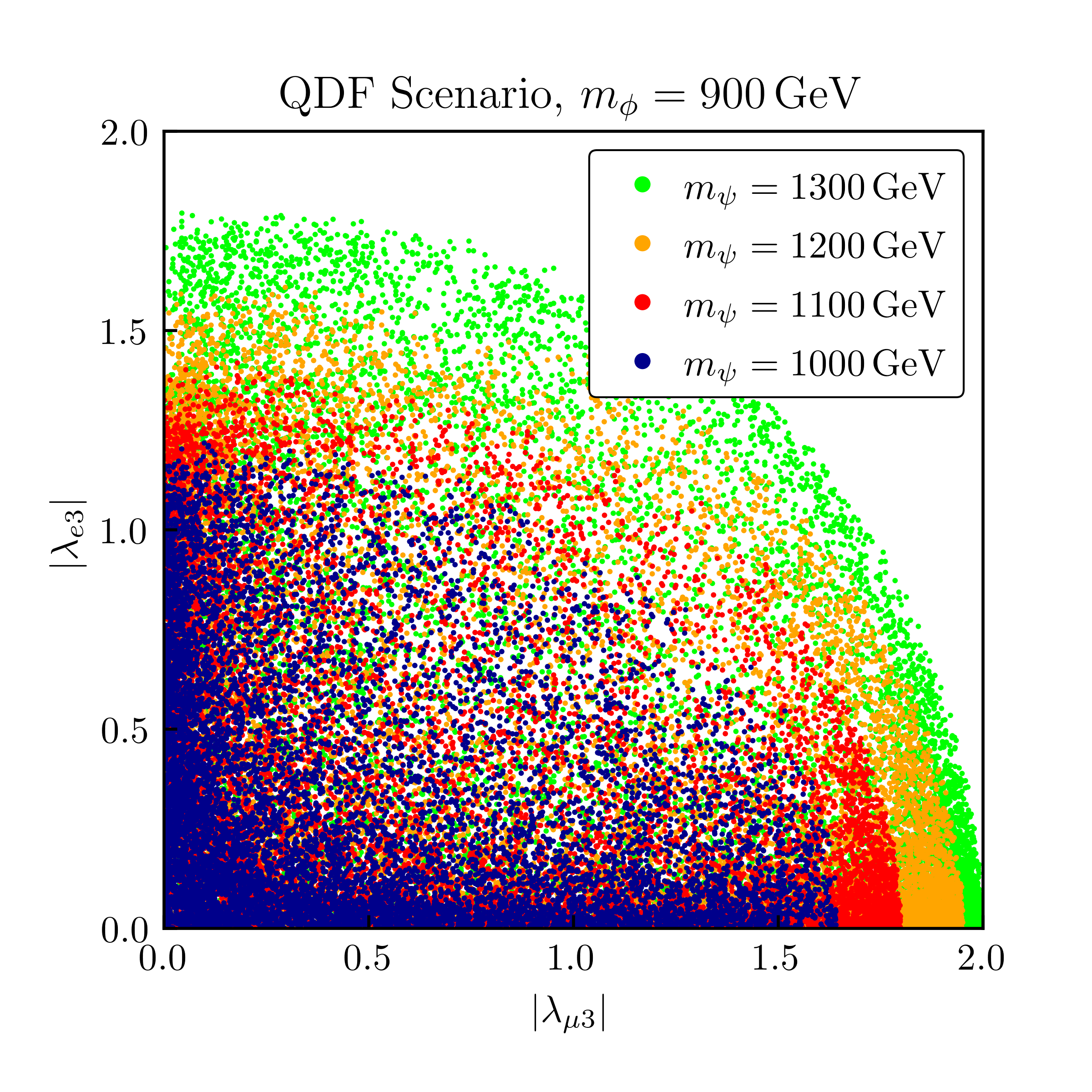}
		\caption{$|\lambda_{\mu 3}| - |\lambda_{e 3}|$ plane}
		\label{fig::cbdqdfb}
		\end{subfigure}
	\caption{Allowed couplings $|\lambda_{i3}|$ for $m_\phi = 900\,\mathrm{GeV}$ and varying $m_\psi$ based on all constraints in the QDF scenario.}
	\label{fig::cbdqdf}
	\end{figure}

After having discussed all the experimental constraints independently in the previous sections, we are now prepared to provide a global picture of our model's viable parameter space. To this end we impose all constraints simultaneously and demand that they are fulfilled at the $2\sigma$ level. 
The results are shown in Figure \ref{fig::cbdmasses}, Figure \ref{fig::cbdsff} and Figure \ref{fig::cbdqdf}.\par   
In Figure \ref{fig::cbdmasses} we show the viable values for the mass parameters $m_\psi$ and $m_{\phi_3}$ in both scenarios. The LHC constraints discussed in Section \ref{sec::collider} are shown as the grey and light-grey exclusion contours. We find that all constraints can only be satisfied simultaneously for mediator masses $m_\psi \GtrSim 750\,\mathrm{GeV}$ and DM masses $m_{\phi_3}\GtrSim 550\,\mathrm{GeV}$. This is due to the combination of direct detection and relic density constraints. As shown in Section \ref{sec::relicdensity}, the observed relic density requires that the couplings $D_i$ are large, since the annihilation rate is $p$-wave suppressed. On the other hand, the direct detection limits force the mediator mass to be large enough in order to suppress $f_\gamma$ sufficiently and compensate for the large couplings $D_i$. For values $750\, \mathrm{GeV} \LessSim m_\psi \LessSim 1200\,\mathrm{GeV}$ this suppression of $f_\gamma$ through $m_\psi$ alone is not sufficient to compensate for sizeable couplings $D_i$. Thus, in this region the combined constraints from direct detection and the relic density additionally demand that the DM mass $m_{\phi_3}$ is comparably large as the XENON1T upper limits on the scattering cross section decrease for increasing DM masses in the region $m_{\phi_3} > 30\,\mathrm{GeV}$. Large values of $m_{\phi_3}$ are also needed, since for such mediator masses the direct detection constraints exclude large couplings $D_i \GtrSim 1.5$ and hence the DM annihilation rate needs to be enhanced through large DM masses in order to yield the correct relic density. For the LHC constraints we find that the largest exclusion in the $m_\psi - m_{\phi_3}$ plane is given in the QDF scenario. As we have chosen the parameter $\eta$ from eq.\@ \eqref{eq::masssplitting} to be negative, the ansatz $D_1=D_2$ necessary for a proper recasting of the LHC limits we consider requires $D_3 > D_{1,2}$ to be consistent with the SFF scenario. Hence, one is always left with a larger coupling to taus than to electrons and muons which results in a reduced branching ratio of the mediator to the latter two. This in turn reduces the signal cross section in the SFF scenario compared to the QDF scenario, where the couplings $D_i$ can be equal. With respect to the overall impact of the LHC constraints we find that the above explained interplay between the direct detection and relic density constraints renders them irrelevant in constraining the parameter space of the model. As far as the mixed-flavour final states $\ell_i \bar{\ell}_j + \slashed{E}_T$ mentioned in Section \ref{sec::collider} are concerned, we find that since only large masses $m_\psi$ and $m_{\phi_3}$ are viable their signal cross section is highly suppressed and yields values of order $\mathcal{O}(\mathrm{ab})$. The SM background, on the other hand, is dominated by the production of $t\bar{t}$ and $WW$ pairs, with significantly larger cross sections than $\psi \bar{\psi}$ pair production in our model: 
we find the background to be of order $\mathcal{O}(\mathrm{pb})$ and thus expect the mixed-flavour final states to not exhibit a significant discovery potential.\par
In Figure \ref{fig::cbdsff} we show the constraints of the combined analysis on the coupling matrix $\lambda$.The structure of the latter is mainly determined by the flavour, relic density and direct detection constraints. The interplay between the former two causes the circular bands shown in Figure \ref{fig::cbdsffa}. In the SFF scenario the relic density constraint reduces to the spherical condition
\begin{equation}
|\lambda_{e3}|^2+|\lambda_{\mu 3}|^2+|\lambda_{\tau 3}|^2 \approx \text{const.}\,,
\end{equation}	
which causes the outer edge of the bands shown in Figure \ref{fig::cbdsffa}. At the same time the flavour physics limit from the LFV decay $\mu \rightarrow e \gamma$ discussed in Section \ref{sec::flavour} forces the product of $|\lambda_{e 3}|$ and $|\lambda_{\mu 3}|$ to be small, which is illustrated in Figure \ref{fig::cbdsffb}, as either $|\lambda_{e 3}|$ can be large while $|\lambda_{\mu 3}|$ is small or vice versa. This in turn causes the inner edge of the bands of Figure \ref{fig::cbdsffa} as well as the points scattered close to the $x$-axis. Figure \ref{fig::cbdsff} also reflects the constraints from direct detection, as the maximum allowed values for $|\lambda_{e 3}|$ are smaller than the ones for $|\lambda_{\mu 3}|$, which can be seen in Figure \ref{fig::cbdsffb}. Also, for large enough values of $|\lambda_{e 3}|$ the DM-nucleon scattering constraints become dominant over the relic density constraint. This explains why in Figure \ref{fig::cbdsffa}  the circular bands become thinner for growing values of $|\lambda_{e 3}|$ above this threshold. For even larger values of $|\lambda_{e 3}|$ we find that the direct detection and relic density constraints cannot be satisfied simultaneously, which strongly disfavours the case of $e$-flavoured DM, i.e.\@ $|\lambda_{e 3}| > |\lambda_{\mu 3}|, |\lambda_{\tau 3}|$. As the DM-nucleon scattering rate is significantly larger for an electron in the loop in Figure \ref{fig::photonpenguin}, we find that even for large masses $m_\psi \GtrSim 1200 \,\mathrm{GeV}$ the DM particle $\phi_3$ needs to be mainly $\mu$- or $\tau$-flavoured in spite of the $1/m_\psi^2$ suppression of $f_\gamma$.\par
While the allowed parameter space of the QDF scenario shown in Figure \ref{fig::cbdqdf} exhibits analogous features as  the SFF scenario, they are far less pronounced. This is mainly due to the different dynamics of the thermal freeze-out, where all dark flavours are present. The relic density constraint thus reduces to the condition
\begin{equation}
\sum_{ij} |\lambda_{ij}|^2 \approx \text{const.}\,,
\end{equation}
i.e.\@ it demands that the couplings $|\lambda_{ij}|$ form the shell of a 9-dimensional sphere. This causes the outer circular edge that can be seen in both Figure \ref{fig::cbdqdfa} and Figure \ref{fig::cbdqdfb}. Since the couplings $D_i$ cannot exhibit a significant splitting in the QDF scenario, the flavour constraints are weaker than in the SFF scenario. This is clearly visible in Figure \ref{fig::cbdqdfb}, where in spite of the higher point density close to the axes points with comparable values for $|\lambda_{e 3}|$ and $|\lambda_{\mu 3}|$ are also allowed. This at the same time explains the absence of an inner edge for the contours of Figure \ref{fig::cbdqdfa}, which again is in contrast to the SFF scenario. The impact of the direct detection limits on the parameter space can again be seen in both panels of Figure \ref{fig::cbdqdf}, as for a mediator mass $m_\psi = 1000\,\mathrm{GeV}$ (blue points), for example, the coupling of DM to tau leptons may become maximal with $|\lambda_{\tau 3}| = 2.0$, while the couplings to muons and electrons are limited to $|\lambda_{\mu 3}| \simeq 1.7$ and $|\lambda_{e 3}| \simeq 1.2$. Since the DM-nucleon scattering rate is suppressed by $1/m_\psi^2$,  the maximum allowed values for $|\lambda_{\mu 3}|$ and $|\lambda_{e 3}|$ increase with a growing mediator mass. In Figure \ref{fig::cbdqdfa} we further see that the direct detection constraint is dominant over the relic density constraint for a larger range of allowed values for $|\lambda_{e 3}|$ than in the SFF scenario. For the DM particle's flavour we find that, while most of the viable parameter points correspond to $\tau$-flavour, a significant part of the parameter space also allows for $\mu$- and $e$-flavoured DM.\par 

The results of the combined analysis also give us an indication which experiments are most likely to discover first traces of our model. Due to the mass scale of the new particles, we do not expect upcoming LHC runs to detect a signal, and future lepton colliders would likely require center-of-mass energies above the TeV scale. The prospects for flavour physics are more optimistic, as the upcoming MEG~II experiment~\cite{Meucci:2022qbh} is expected to strengthen the upper limits on $\mu \to e \gamma$ by almost one order of magnitude. Concerning DM detection experiments, the viable mass range for DM favours a discovery in future direct detection experiments like XENONnT \cite{XENON:2020kmp} and DARWIN \cite{DARWIN:2016hyl} over an observation in indirect detection experiments, for which we found relevant constraints mostly in the low-mass region.
\section{Summary and Outlook}

In this paper we studied a simplified model of lepton-flavoured dark matter beyond MFV, as introduced in Section \ref{sec::theory}. In this model, the SM is extended  by a DM flavour triplet $\phi$, which is a complex scalar and a singlet under the SM gauge group, and it transforms as triplet under a new dark flavour symmetry $U(3)_\phi$. The DM field $\phi$  couples to the right-handed charged leptons $\ell_R$ via a charged fermionic mediator $\psi$, carrying the same gauge quantum numbers as $\ell_R$. The interaction between the SM leptons and DM is governed by the coupling $\lambda$, a generic $3 \times 3$ complex matrix. Following the DMFV ansatz we assumed that $\lambda$ and the SM Yukawa couplings are the only sources that violate the extended flavour symmetry of the model. 
In order to study the phenomenology of this model and constrain its parameter space we then explored the limits from collider searches, LFV decays, the observed DM relic density, direct detection and indirect detection experiments.\par
In Section \ref{sec::collider} we found that the most stringent limits in terms of LHC searches stem from SUSY searches for sleptons of the first and second generation that we recasted for mediator pair-produtcion in our model. Here we used a CMS search for the final state $\ell\bar{\ell}+ \slashed{E}_T$ based on the run 2 LHC data with an integrated luminosity of $137\,\mathrm{fb}^{-1}$. The exclusion in the mass plane $m_\psi - m_\phi$ was found to be the largest for vanishing couplings $D_3$ and large couplings $D_1 = D_2$. For large couplings $D_3$, the reduced branching ratio of the mediator decay to final states with an electron or a muon  in turn also lowers the relevant signal cross section. We concluded that current LHC limits maximally exclude masses $m_\psi \LessSim 750\,\mathrm{GeV}$.\par
To study the flavour structure of $\lambda$, in Section \ref{sec::flavour} we investigated the constraints that LFV decays put on our model. The strongest limit stems from the decay $\mu \rightarrow e \gamma$ and forces the coupling matrix $\lambda$ to satisfy $|\left(\lambda\lambda^\dagger\right)_{\mu e}|^{1/2} \LessSim m_\psi/15\,\mathrm{TeV}$ for masses $m_\psi \gg m_{\phi_3}$. We further found that in the limit of vanishing flavour mixing angles $\theta_{ij}$ or degenerate couplings $D_i$ the LFV decay constraints are rendered irrelevant. In Section \ref{sec::flavour} we also commented on constraints from precision measurements of leptonic MDMs and EDMs. These constraints are only relevant at new physics scales of order $\mathcal{O}(100 \,\mathrm{GeV})$, which are already excluded by LHC searches.\par 
For the analysis of the constraints the observed DM relic abundance places on our model we have defined two benchmark scenarios for the thermal freeze-out of DM in Section \ref{sec::relicdensity}. The first one was the QDF scenario, where we assumed a negligible mass splitting between the different DM generations, such that all dark flavours are present during freeze-out. In the SFF scenario on the other hand, we assumed that the mass splitting is significant, such that only $\phi_3$ contributes to the latter. We found that in both scenarios the thermally averaged annihilation cross section is $p$-wave suppressed. In turn this forces the couplings $D_i$ to be large in order not to yield a too small annihilation rate or, equivalently, overabundant DM.\par
In Section \ref{sec::directdetection} we  discussed the  phenomenology of direct detection experiments. After identifying relevant interactions for scatterings between DM and SM matter in our model, we found that the dominant process for direct detection is DM-nucleon scattering induced by a one-loop photon penguin diagram. Using XENON1T data we then constrained the parameter space of our model, finding that the smaller the mass of the lepton in the loop, the larger the contribution to the DM-nucleon scattering cross section.  Hence, the most stringent limits are placed on the coupling $|\lambda_{e 3}|$, which in turn disfavours $e$-flavoured DM.\par 
Section \ref{sec::indirectdetection} was dedicated to the analysis of indirect detection constraints. Since the leading annihilation rate of DM into a pair of SM leptons exhibits a $p$-wave suppression, we included the subleading one-loop annihilation into two photons and annihilation into the three-body final state with two leptons and a photon. We found that the indirect detection constraints are generally weak, and only the constraints from measurements of the $\gamma$-ray spectrum yield relevant limits for mediator masses up to $m_\psi \simeq 1000\,\mathrm{GeV}$ in the near-degeneracy region $m_\psi \approx m_{\phi_3}$.\par 
To provide a global picture we then used Section \ref{sec::combined} to perform a combined analysis of all constraints. Here we found that the combination of relic density and direct detection constraints renders LHC limits irrelevant and forces the masses to roughly lie in the ranges $750\,\mathrm{GeV} \LessSim m_\psi \LessSim 1550\,\mathrm{GeV}$ and $550\,\mathrm{GeV} \LessSim m_{\phi_3} < m_\psi$. Further, the allowed values for the couplings $|\lambda_{i3}|$ are mainly determined by the interplay of relic density, flavour and direct detection constraints in both freeze-out scenarios. 
Last but not least, we determined the flavour of DM and found that while in the SFF scenario only $\mu$- and $\tau$-flavoured DM are viable, the QDF scenario also allows for $e$-flavoured DM.\par 
We conclude that lepton-flavoured complex scalar DM is a viable DM candidate in the context of current experimental constraints. Thanks to its non-trivial flavour structure allowed within the DMFV framework, it is governed by a rich phenomenology in which the synergy of various experiments yields important information on the structure of the model. With future improved sensitivities, we may hope to discover first hints of lepton-flavoured DM in the laboratory.
 
\paragraph*{Acknowledgements}
This work is supported by the Deutsche Forschungsgemeinschaft (DFG, German Research Foundation) under grant 396021762 -- TRR 257. P.A. is supported by the STFC under Grant No. ST/T000864/1. H.A.\@ acknowledges the scholarship and support he receives from the Avicenna-Studienwerk e.V., the support of the doctoral school ``Karlsruhe School of Elementary and Astroparticle Physics: Science and Technology (KSETA)'' and the funding he received for his academic visit at the University of Oxford from the ``Karlsruhe House of Young Scientists (KHYS)''. He further acknowledges the hospitality during his academic visit at the Rudolph Peierls Centre for Theoretical Physics. M.B. thanks the Mainz Institute for Theoretical Physics (MITP) of the Cluster of Excellence PRISMA$^+$ (Project ID 39083149) for its hospitality and support during the completion of this project.

\appendix
\section{Partial Wave Expansion Coefficients}
\label{app::partialwave}
The coefficients from the partial wave expansion of the thermal averaged annihilation cross section from eq.\@ \eqref{eq::partialwave} with the full final state mass dependence read 
\begin{align}
\nonumber
a &= \sum_{ijkl} |\lambda_{ik}|^2|\lambda_{jl}|^2\frac{\sqrt{-2 m_{\ell _k}^2 \left(m_{\ell _l}^2+4 m_{\phi }^2\right)+m_{\ell _k}^4+\left(m_{\ell _l}^2-4 m_{\phi }^2\right){}^2}}{128 \pi  m_{\phi }^4 \left(m_{\ell _k}^2+m_{\ell _l}^2-2 m_{\psi }^2-2 m_{\phi }^2\right){}^2}\\
&\phantom{=}\times\left(4 m_{\phi }^2 \left(m_{\ell _k}^2+m_{\ell _l}^2\right)-6 m_{\ell _k}^2
   m_{\ell _l}^2+m_{\ell _k}^4+m_{\ell _l}^4\right)\,,\\
   \nonumber
   \phantom{=}&\\
\nonumber
b &= \sum_{ijkl} |\lambda_{ik}|^2|\lambda_{jl}|^2\frac{\sqrt{-2 m_{\ell _k}^2 \left(m_{\ell _l}^2+4 m_{\phi }^2\right)+m_{\ell _k}^4+\left(m_{\ell _l}^2-4 m_{\phi }^2\right){}^2}}{3072 \pi  m_{\phi }^4 \left(m_{\ell _k}^2+m_{\ell _l}^2-2 m_{\psi }^2-2 m_{\phi }^2\right){}^4}\\
\nonumber
&\phantom{=}\times \biggl\{3 \left(-4 m_{\phi }^2 \left(m_{\ell _k}^2+m_{\ell _l}^2\right)+6 m_{\ell
   _k}^2 m_{\ell _l}^2+m_{\ell _k}^4+m_{\ell _l}^4\right) \left(m_{\ell _k}^2+m_{\ell _l}^2-2 m_{\psi }^2-2 m_{\phi }^2\right){}^2\\
   \nonumber
   &\phantom{=} -\frac{3 \left(-16 m_{\phi }^2 \left(m_{\ell _k}^2+m_{\ell _l}^2\right)+3
   \left(m_{\ell _k}^2-m_{\ell _l}^2\right){}^2+16 m_{\phi }^4\right)}{-2 m_{\ell _l}^2 \left(m_{\ell _k}^2+4 m_{\phi }^2\right)+\left(m_{\ell _k}^2-4 m_{\phi }^2\right){}^2+m_{\ell _l}^4}\\
   \nonumber
    &\phantom{=}\times\left(4 m_{\phi }^2 \left(m_{\ell _k}^2+m_{\ell _l}^2\right)-6 m_{\ell _k}^2 m_{\ell _l}^2-m_{\ell _k}^4-m_{\ell _l}^4\right) \left(m_{\ell
   _k}^2+m_{\ell _l}^2-2 m_{\psi }^2-2 m_{\phi }^2\right){}^2 \\
   \nonumber
   &\phantom{=}+8 \biggl[-\left(m_{\ell
   _k}^2-m_{\ell _l}^2\right){}^2 \left(m_{\psi }^2 \left(m_{\ell _k}^2+m_{\ell _l}^2\right)+3 m_{\ell _k}^2 m_{\ell _l}^2+m_{\psi }^4\right)\\
   \nonumber
   &\phantom{=}+8 m_{\phi }^6 \left(8 m_{\psi }^2-7 \left(m_{\ell _k}^2+m_{\ell
   _l}^2\right)\right)\\
   \nonumber
   &\phantom{=}+m_{\phi }^4 \left(-96 m_{\psi }^2 \left(m_{\ell _k}^2+m_{\ell _l}^2\right)+66 m_{\ell _k}^2 m_{\ell _l}^2+31 m_{\ell _k}^4+31 m_{\ell _l}^4+32 m_{\psi }^4\right)\\
   \nonumber
   &\phantom{=}+2 m_{\phi }^2 \biggl(4 m_{\psi
   }^4 \left(m_{\ell _k}^2+m_{\ell _l}^2\right)+m_{\psi }^2 \left(46 m_{\ell _k}^2 m_{\ell _l}^2+9 m_{\ell _k}^4+9 m_{\ell _l}^4\right)\\
  &\phantom{=}-2 \left(m_{\ell _k}^2+m_{\ell _l}^2\right) \left(4 m_{\ell _k}^2 m_{\ell
   _l}^2+m_{\ell _k}^4+m_{\ell _l}^4\right)\biggr)+32 m_{\phi }^8\biggr]\Biggr\}\,.
\end{align}
\bibliography{ref}
\bibliographystyle{JHEP}
\end{document}